# Gosling Designer: a Platform to Democratize Construction and Sharing of Genomics Data Visualization Tools

**Sehi L'Yi[1], John Conroy[1], Priya Misner[1], David Kouřil[1], Astrid van den Brandt[1], Lisa Choy[1], Nezar Abdennur[2], Nils Gehlenborg[1]**

[1] Harvard Medical School, Boston, MA, USA [2] UMass Chan Medical School, Worcester, MA, USA

* Corresponding author: nils@hms.harvard.edu

Analysis of genomics data is central to nearly all areas of modern biology. Despite significant progress in artificial intelligence (AI) and computational methods, these technologies require significant human oversight to generate novel and reliable biological insights. Consequently, the genomics community has developed a substantial number of diverse visualization approaches[1] and a proliferation of tools that biologists rely on in their data analysis workflows[2]. While there are a few commonly used visualization tools for genomics data, such as IGV[3], Circos[4], or JBrowse[5], many tools target specific use cases for genomics data interpretation and offer only a limited, predefined set of visualization types. Moreover, static visualizations—which are the output of the majority of existing tools—often fail to support exploratory analysis. Developing interactive visualizations and tools typically requires significant time and technical expertise, even when supported by modern LLM-powered coding assistants, and the resulting visualizations can be difficult to share among collaborators.

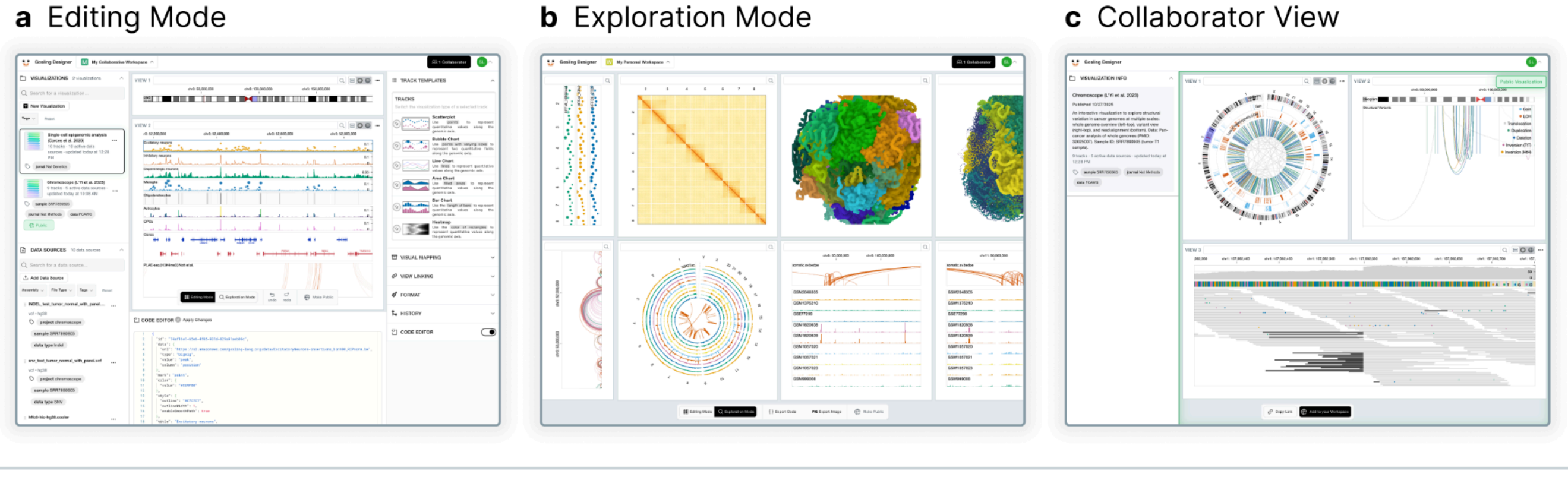

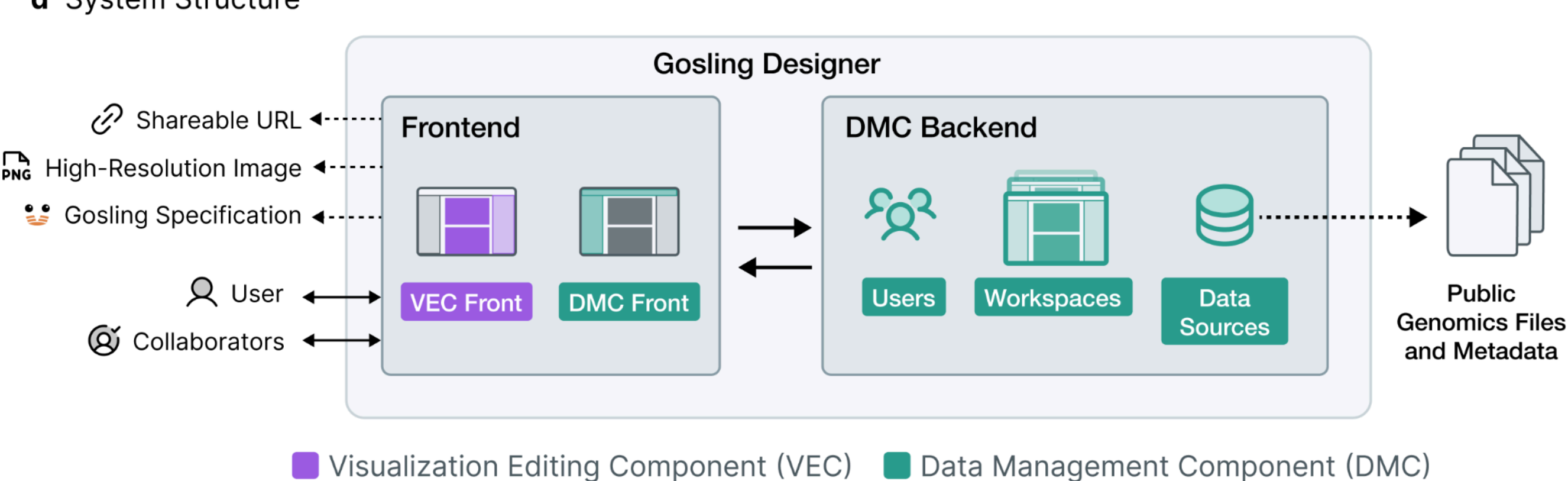

**Figure 1. Overview of the user interfaces and system structure of Gosling Designer. a**, The interface in *editing* mode with genome browser tracks. **b**, The interface in *exploring* mode, displaying a genome interaction matrix with a 3D genome view. **c**, A Chromoscope[6] visualization created using Gosling Designer that is shared with a collaborator. **d**, The schematic diagram showing the backend-frontend model of Gosling Designer with the visualization editing component and the data management component.

We developed Gosling Designer, an all-in-one platform for editing, exploring, and sharing visualizations of genomics data (Figure 1). Gosling Designer aims to support a broad range of users in genomics, regardless of programming and visualization expertise, such as clinicians, experimentalists, and computational biologists, to use interactive visualizations in their workflows and share them with collaborators for interpretation. Gosling Designer addresses four key challenges observed in existing genomics visualization tools: (1) limited versatility, (2) difficulty of visualization authoring, (3) complexity of data management, and (4) barriers to sharing and collaboration. Gosling Designer is available via https://designer.gosling-lang.org.

Gosling Designer builds on the Gosling grammar[7], enabling the creation of a broad spectrum of genome-mapped data visualizations tailored to diverse analysis use cases. It supports not only the creation of conventional genomics visualizations but also the design of novel depictions driven by emerging genomics technologies. In addition to visualizing genome sequences in two-dimensional (2D) space using linear and circular layouts, Gosling Designer also allows three-dimensional (3D) representations, such as the physical structure of genomes. Through this expressive power of Gosling Designer, users can create classic genome browser tracks, ideograms, genome interaction matrices, interactive Circos-like plots[4], and 3D genome structures. Visualizations created in Gosling Designer incorporate powerful built-in user interactions, including seamless zooming and panning, and support linked multiple views for effective analysis. This further enables combining views for effective data exploration, such as a Chromoscope[6] visualization that combines a Circos plot as a whole genome overview with linked genome browser tracks for showing narrow genomic regions across scales. Example visualizations are provided in the Supplemental Notes.

Gosling Designer democratizes genomics data visualization by providing a graphical interface that enables a broad range of genomics stakeholders—including experimentalists, clinicians, computational biologists, and software developers—to create customized visualizations with ease. Through familiar user interactions, such as drag and drop, users can construct complex multi-view visualizations without programming. It offers an extensive set of starter visualization examples and track templates that allow users to quickly adapt designs to their own data or switch between alternative visualization options within a gallery. Gosling Designer's flexible visual mapping, a paradigm familiar from general-purpose visualization tools, allows fine-grained control over how visual representations are designed (e.g., selecting mark types or adjusting visual channels). For advanced users, an integrated code editor exposes the underlying Gosling grammar, providing full programmatic control. The aforementioned user interfaces—templates, visual mapping, and code editor—are linked bidirectionally, lowering users' learning curve and facilitating iterative design, as demonstrated in prior visualization research[8]. Finally, Gosling Designer records interaction histories and presents them in an interactive view, enabling users to easily explore, undo, or revisit previous visualization states.

Gosling Designer includes an integrated data management platform that streamlines the organization and reuse of genomics data sources for visualization. Users can create accounts to organize genomics files in their workspaces. Users can create personal or shared workspaces to add and manage references to publicly accessible genomics data source, e.g., in community data repositories and associated metadata, such as genome assemblies and assay types. This data structure facilitates efficient data discovery through metadata- or keyword-based filtering, allowing users to locate datasets of interest for visualization. Gosling Designer supports a broad range of standard genomics file formats (e.g., BigWig, BED, GFF, BAM, VCF, CSV/TSV), enabling seamless visualization without the need for additional data preprocessing. See the Supplemental Notes for the full list of supported file formats.

Gosling Designer supports collaborative construction of genomics data visualizations—a critical need in genomics analysis that is often unmet by existing tools[2]. Users can invite collaborators to shared workspaces with configurable access levels (viewer, editor, or administrator), allowing joint management and editing of visualizations and data sources. For streamlined sharing, users can generate private links that are viewable by shared users only, providing direct access to previously created visualizations. All visualizations are fully compatible with the broader Gosling ecosystem, enabling exported specifications to be seamlessly reused in Python notebooks through Gos[9] or integrated into interactive web applications via Gosling.js[7,9].

By combining flexibility, usability, and collaboration, Gosling Designer democratizes the creation of both established and novel genome-mapped data visualizations. This broadens participation in visualization design and empowers diverse users to more effectively interpret, analyze, and communicate complex genomics data. The use of the Gosling grammar provides flexibility in extending Gosling Designer for additional features. We envision that the Gosling Designer platform will play an important role in efficient human-AI collaboration for analysis of genomic data.

## Acknowledgements

This work was supported by the National Institutes of Health (R01HG011773, K99HG013348, U24OD038421). The project was inspired by the Reservoir Genomics Platform (https://resgen.io) created by Peter Kerpedjiev and Nezar Abdennur.

## References

1. Nusrat, S., Harbig, T. & Gehlenborg, N. Tasks, Techniques, and Tools for Genomic Data Visualization. *Comput. Graph. Forum* **38**, 781–805 (2019).

2. van den Brandt, A., L'Yi, S., Nguyen, H. N., Vilanova, A. & Gehlenborg, N. Understanding visualization authoring techniques for genomics data in the context of personas and tasks. *IEEE Trans. Vis. Comput. Graph.* **31**, 1180–1190 (2025).

3. Robinson, J. T., Thorvaldsdottir, H., Turner, D. & Mesirov, J. P. igv.js: an embeddable JavaScript implementation of the Integrative Genomics Viewer (IGV). *Bioinformatics* **39**, (2023).

4. Krzywinski, M. *et al.* Circos: an information aesthetic for comparative genomics. *Genome Res.* **19**, 1639–1645 (2009).

5. Diesh, C. *et al.* JBrowse 2: a modular genome browser with views of synteny and structural variation. *Genome Biol.* **24**, 74 (2023).

6. L'Yi, S. *et al.* Chromoscope: interactive multiscale visualization for structural variation in human genomes. *Nat. Methods* **20**, 1834–1835 (2023).

7. L'Yi, S., Wang, Q., Lekschas, F. & Gehlenborg, N. Gosling: A grammar-based toolkit for scalable and interactive genomics data visualization. *IEEE Trans. Vis. Comput. Graph.* **28**, 140–150 (2022).

8. L'Yi, S., van den Brandt, A., Adams, E., Nguyen, H. N. & Gehlenborg, N. Learnable and expressive visualization authoring through blended interfaces. *IEEE Trans. Vis. Comput. Graph.* 1–11 (2024) doi:10.1109/TVCG.2024.3456598.

9. Manz, T., L'Yi, S. & Gehlenborg, N. Gos: a declarative library for interactive genomics visualization in Python. *Bioinformatics* **39**, (2023).

# Supplemental Notes

## 1. Definitions: Tracks and Views

Gosling Designer distinguishes two important concepts for genomics data visualization: **tracks** and **views**.

- A **track** refers to a unit visualization that typically visualizes a single data source. This term is commonly used in genome browsers. A track can visualize a data source in different ways, using different visualization types, e.g., line charts, bar charts, scatterplots, and heatmaps.
- A **view** refers to a set of tracks that are placed together and shows the identical genomic region. The tracks that belong to the same view are linked in a way that whenever the user zooms and pans, they always show the same genomic region, helping the user to see correlation of multiple datasets.

We use these terms consistently throughout the manuscript and in the user interfaces of the Gosling Designer.

## 2. System Structure

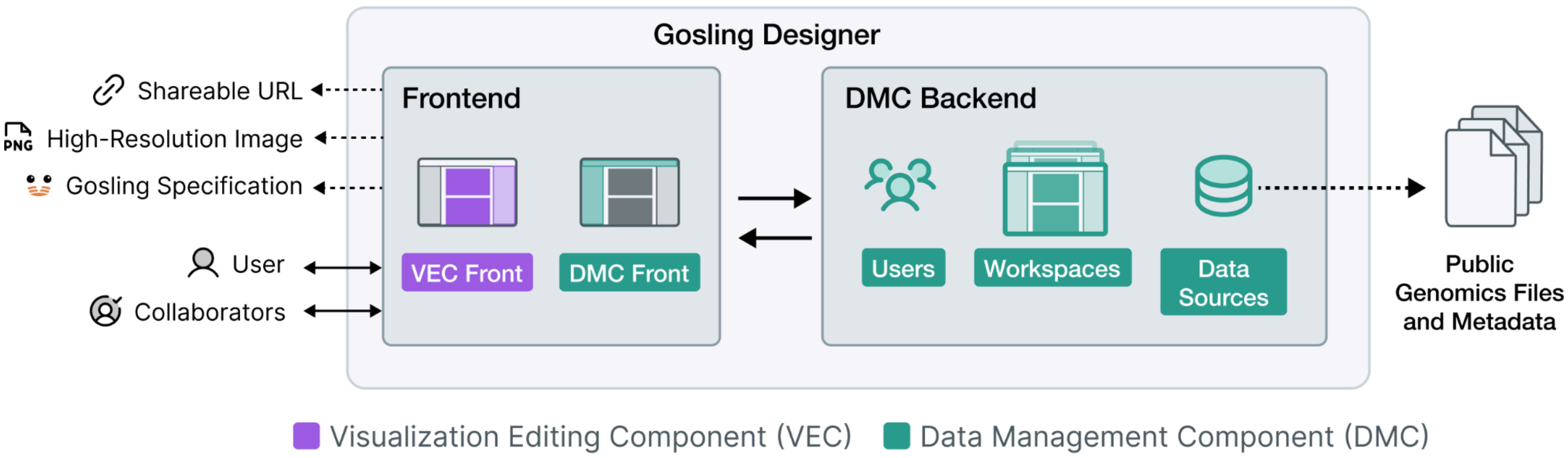

This schematic diagram shows the backend-frontend model of Gosling Designer with two major components: **Visualization Editing Component (VEC)** and **Data Management Component (DMC)**.

- The **VEC** handles user interactions related to editing and exploration of genomics data visualization. The VEC uses Gosling.js[1] to display interactive genomics data visualization in a scalable manner. Upon user interactions taken in the VEC, the underlying Gosling specification (JSON) is manipulated by VEC, which results in updating the visualization displayed.
- The **DMC** manages information about users, workspaces, and data sources (including their public URLs and metadata). This allows the user to add data sources, workspaces, and collaborators to their accounts.

## 3. Supported File Formats

The table below summarizes supported file formats in Gosling Designer. A few file formats that require a HiGlass[2] server are denoted in the corresponding column where ✔ indicates that the corresponding file requires the server. We plan to support additional standard genomics file formats in the future.

| **File Format** | Requires HiGlass Server | Notes |
|---|---|---|
| BAM | – | An index file is required (i.e., BAI). |
| BED | – | Supports BED3, BED6 and BED12. An index file is required (i.e., Tabix). If an index file is missing, the BED file can be added through CSV/TSV file type, with possible performance limitations for large files. |
| BigWig | – | Can be used without an index file. |
| Cooler | – | Can be used without an index file. |
| CSV/TSV | – | Can also be used to upload genomics specific delimited files, e.g., BED, and BEDPE. |
| GFF3 | – | An index file is required (i.e., Tabix). |
| VCF | – | An index file is required (i.e., Tabix). |
| BEDDB | ✔ | HiGlass-specific format for preaggregated BED data. |
| Vector | ✔ | HiGlass-specific format for 1D quantitative data along genomic position, e.g. BigWig. |
| Multi-vec | ✔ | HiGlass-specific format for 2D vector data, one dimension being genomic position, e.g., multiple BigWig files combined. |

## 4. Reusing Specifications for Genomics Workflows

The Gosling specification (JSON) exported from Gosling Designer can be used in the various Gosling[1] ecosystems, such as in Python and JavaScript environments. In addition, Gosling specifications created using external tools (e.g., in the Gos[3] Python Package) can be imported to Gosling Designer for further editing of the visualization. Such functionalities of Gosling Designer enable users to more easily adopt interactive visualizations in their genomics workflows.

### Python Environment

The exported specification can be used in Python notebooks via Gos[3] (https://gosling-lang.github.io/gos/). The supported notebooks include Jupyter Notebooks, Jupyter Labs, and Google Colab. The minimal steps to reuse visualizations in the notebooks are the following.

First, install the Gos package (e.g., in the terminal).

```
1 pip install 'gosling[all]'
```

Then, in a computational notebook, import the package and use the specification with the Gos package.

```
1 import gosling as gos
2 gos.View(exported_specification) # spec in the Python dictionary
```

This will display the interactive Gosling visualization as the output of the code block.

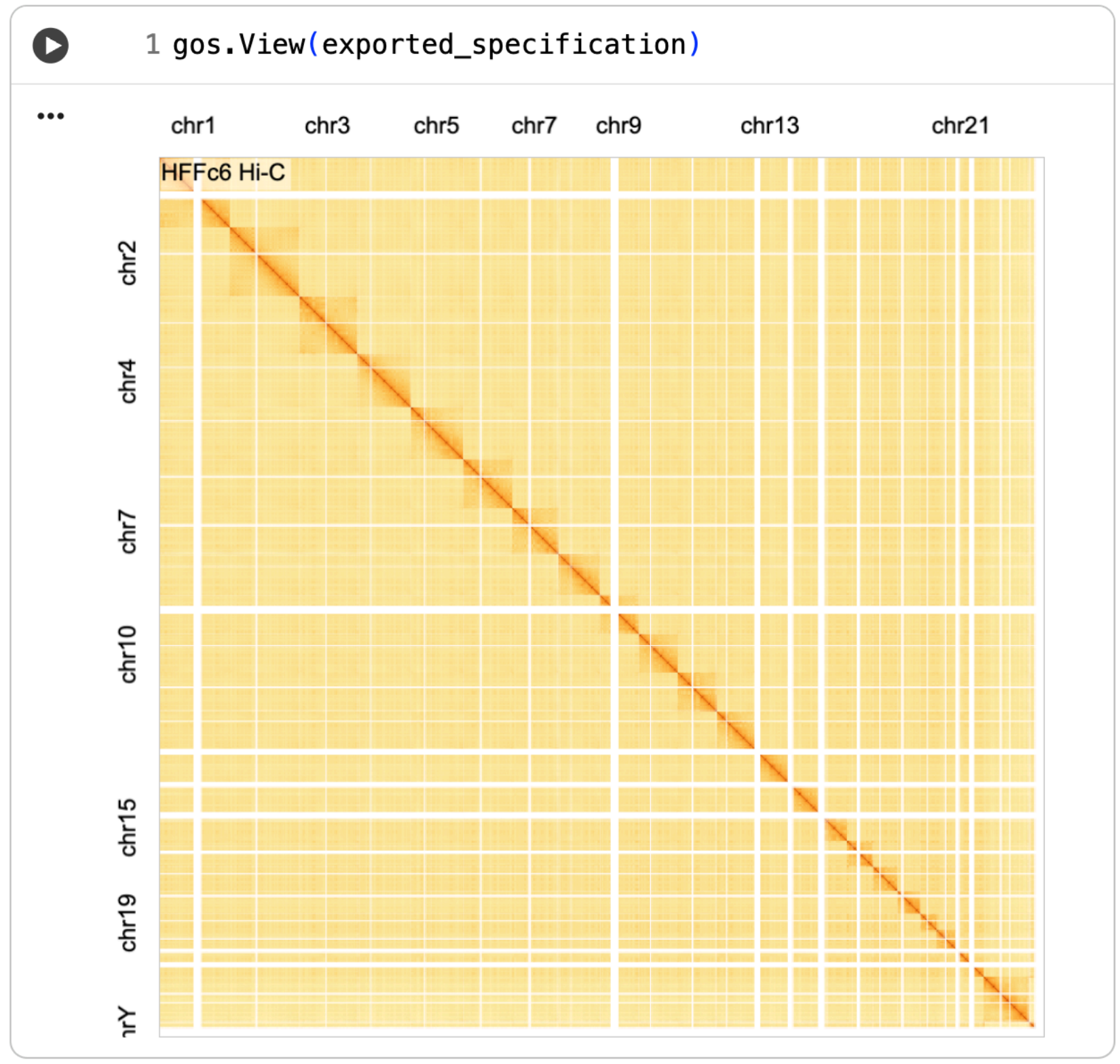

General instructions on using Gos are described in the official documentation (https://gosling-lang.github.io/gos/).

## Web Environment (JavaScript/TypeScript)

The exported specification can be used to build a website that contains the interactive Gosling visualization. This can be useful for making interactive websites (e.g., visualization tools, data portals) that contain visualizations for genomics data with other useful components (e.g., data tables that are linked with visualizations).

For this use case, users can use the specification in the Gosling.js[1] JavaScript library (https://gosling-lang.org/).

```
1 import { GoslingComponent } from "gosling.js";
2
3 function App() {
4   return <GoslingComponent spec={exported_specification} />;
5 }
```

This requires adding Gosling and other dependencies to the users' project.

```
1 npm install gosling.js@alpha react react-dom pixi.js
```

Detailed instructions on using Gosling.js are described in the official documentation (https://gosling-lang.org/).

## Online Editor

The exported specification can be directly used in the Gosling online editor via https://gosling.js.org/ or https://v2.gosling-lang.org/ for using the upcoming version of Gosling that supports 3D visualizations. This editor can be used directly via the browser and does not require installing anything. This helps to further edit the specification together with an interactive visualization rendered on the side. Also, the editor enables browsing a data table for each data source (e.g., available columns and their values).

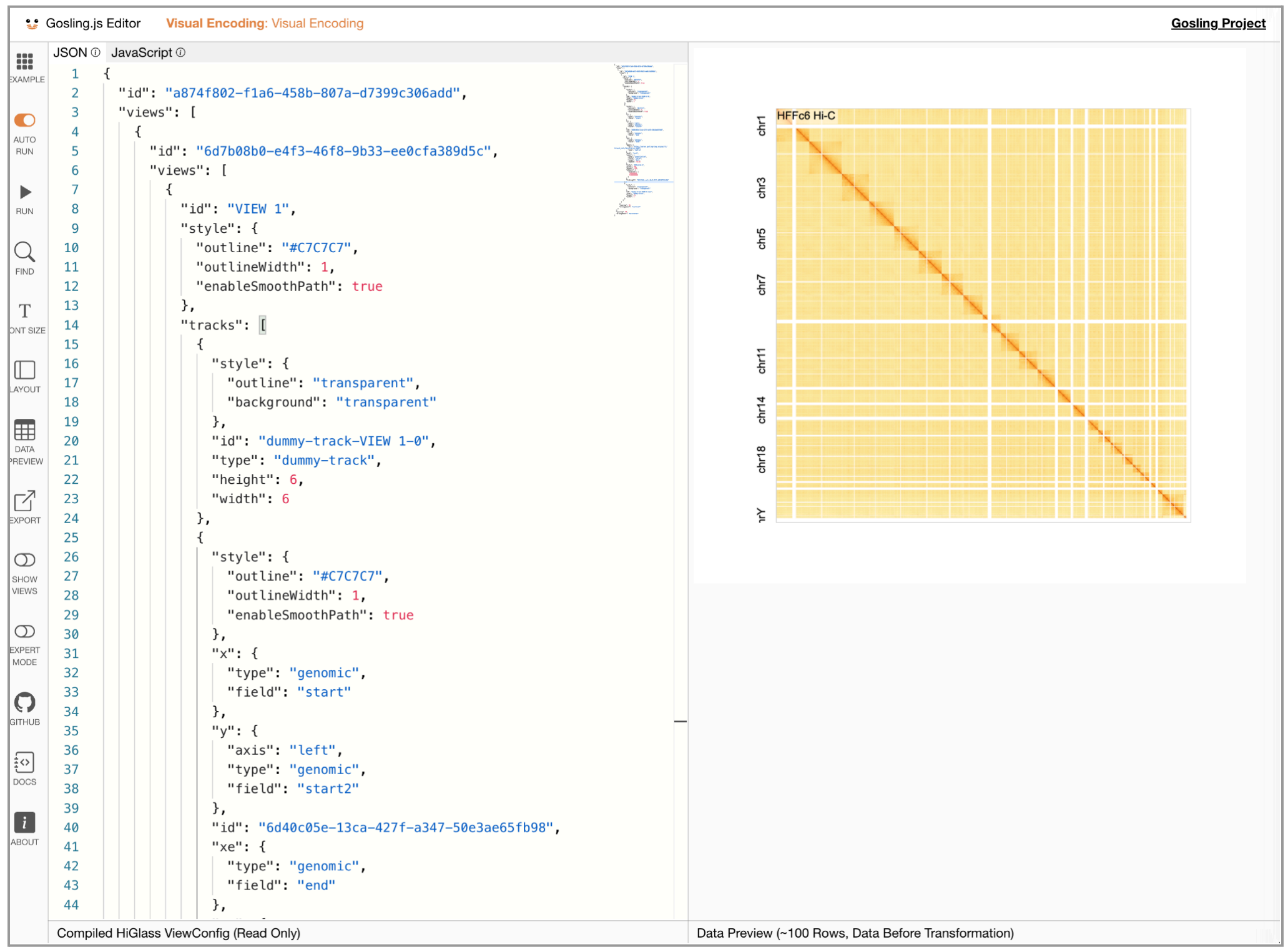

## Reusing Specifications in Gosling Designer

Gosling specifications can be used directly in Gosling Designer. This enables making changes to the visualization with features supported in Gosling Designer. There are many visualization examples for Gosling that can be reused in Gosling Designer, such as in its documentation website (https://gosling-lang.org) and Chromoscope[4] (https://chromoscope.bio/).

As an illustrative example, we describe how Chromoscope visualization can be reused in Gosling Designer. Chromoscope is a multiscal visualization tool for structural variation of cancer genomes. Its visualization is created using Gosling, and the website supports exporting the Gosling specifications.

On the Chromoscope website, the user can download the specification by clicking on the JSON icon in the header.

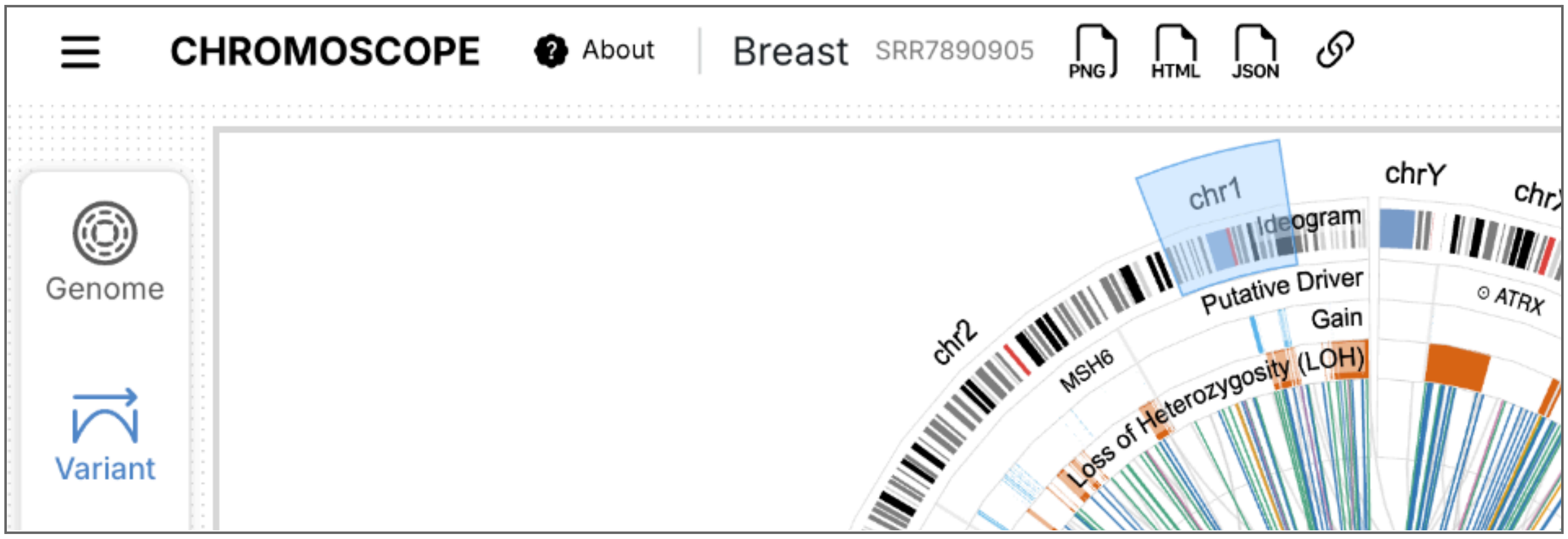

This specification can then be directly used in the Code Editor of Gosling Designer by copying and pasting the code and clicking on the Apply Changes.

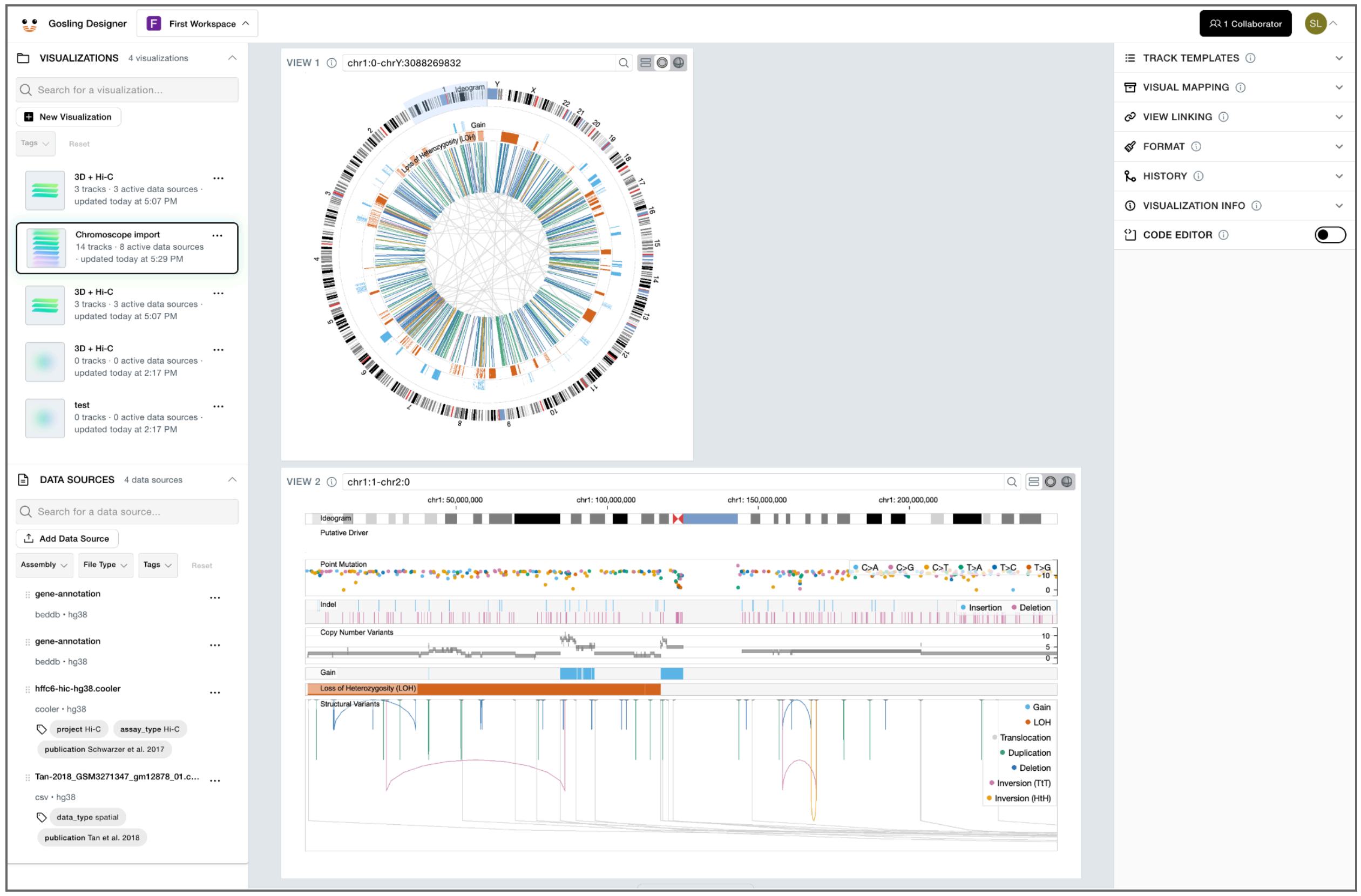

## 5. The Gosling Designer User Interfaces

### Landing Page

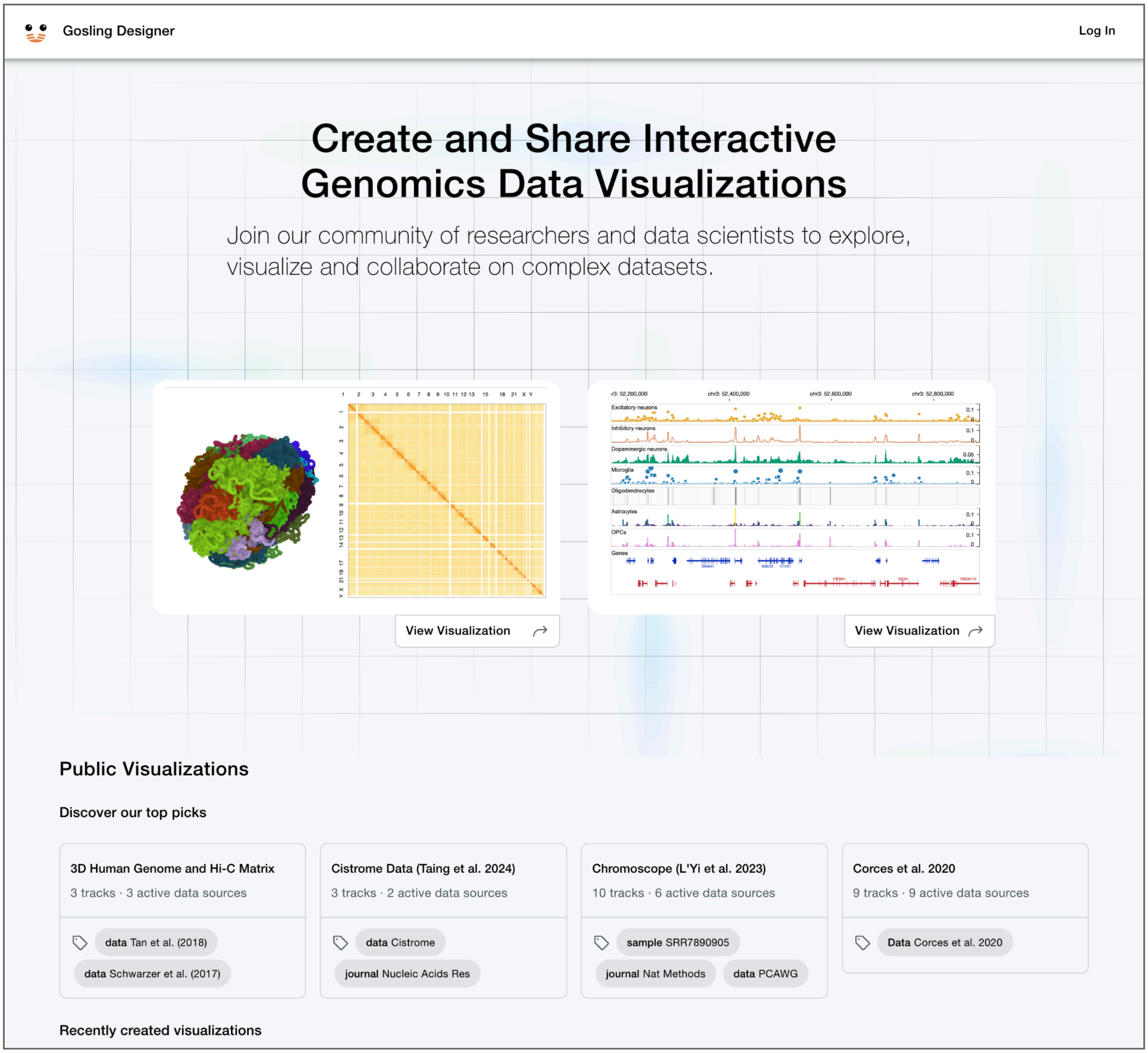

The landing page showcases public visualizations that were created by the Gosling Designer community. The user can add such visualizations to their workspace for further editing and exploration. The new user can also create an account and log in to create their personal workspace from this page.

## Login Page

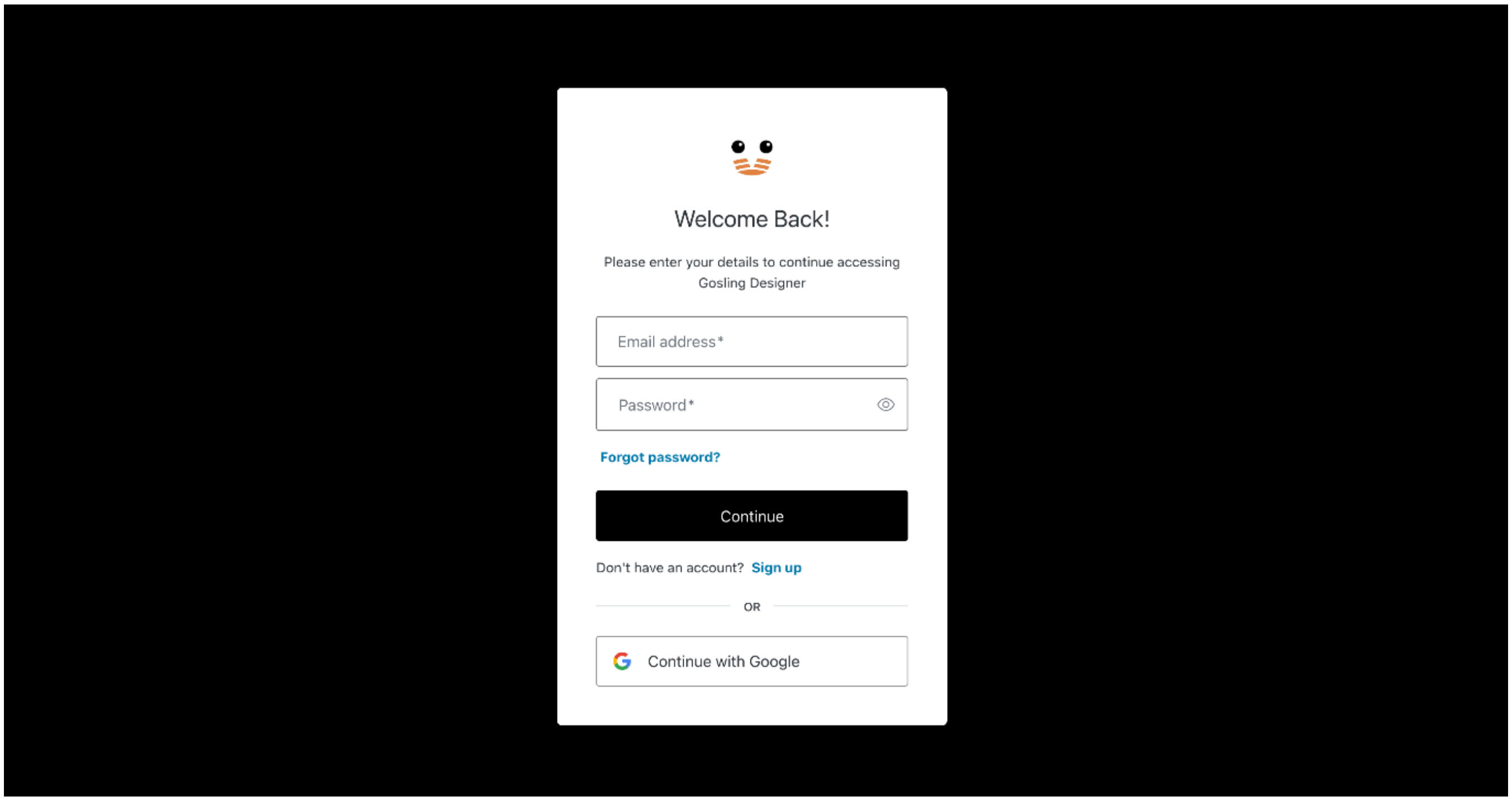

From the landing page, the user can click a login button which redirects to the login page. Here, the user can create an account using the combination of an email and a password or login using their existing account.

## Main User Interface

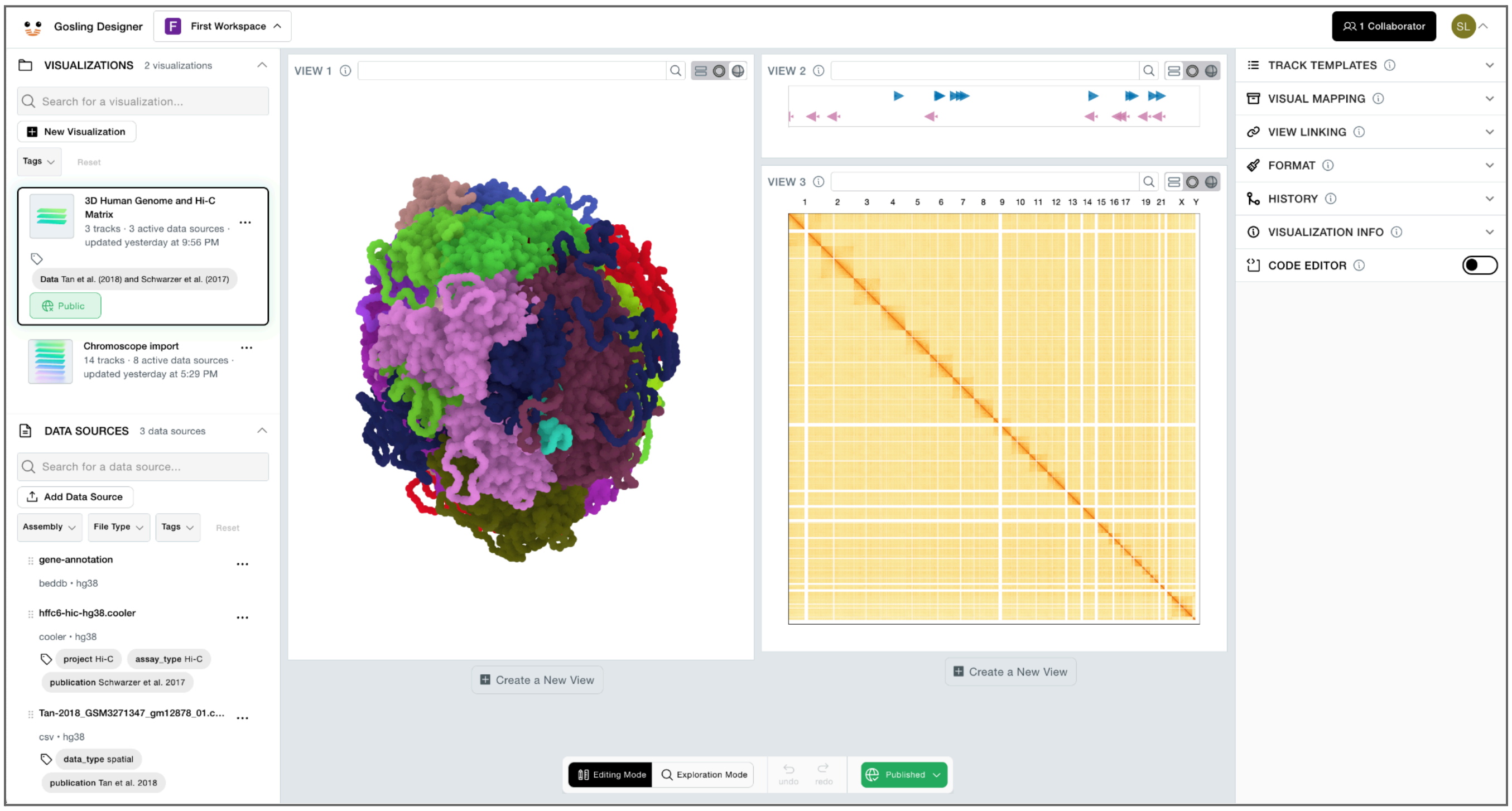

The main user interface of Gosling Designer consists of three main panels:

1. The **Visualizations and Data Sources** panel (top-left): The panel shows information of the visualizations and data sources added in the past. Users can add additional visualizations and data sources in this panel.
2. The **Visualization Canvas** (center): The panel displays the interactive visualization and shows a button group at the bottom that supports options for editing, exploring, and sharing visualizations, such as switching between the editing and exploration modes, exporting specifications and images, undoing and redoing changes to the visualizations, and making visualizations public.
    a. Data sources can be dragged and dropped to this canvas to add a track using the given data source (e.g., dropping on the existing track, or above or below an existing track). The user can create a new view by clicking buttons located on the right or bottom of existing views.
    b. When the visualization becomes public, it will be shown in the landing page where everyone can explore and edit the visualization in their workspaces.
3. The **Visualization Editing** panel (right): The panel supports seven collapsible subpanels for editing visualizations, including Track Templates, Visual Mapping, View Linking, Format, History, Visualization Info, and Code Editor. These subpanels are described separately later in the section. When the user switches from the Editing Mode to Exploration Mode, this panel disappears, allowing users to interact with the visualization (e.g., zoom and pan) with a bigger screen space.

## Main User Interface for Public Visualization

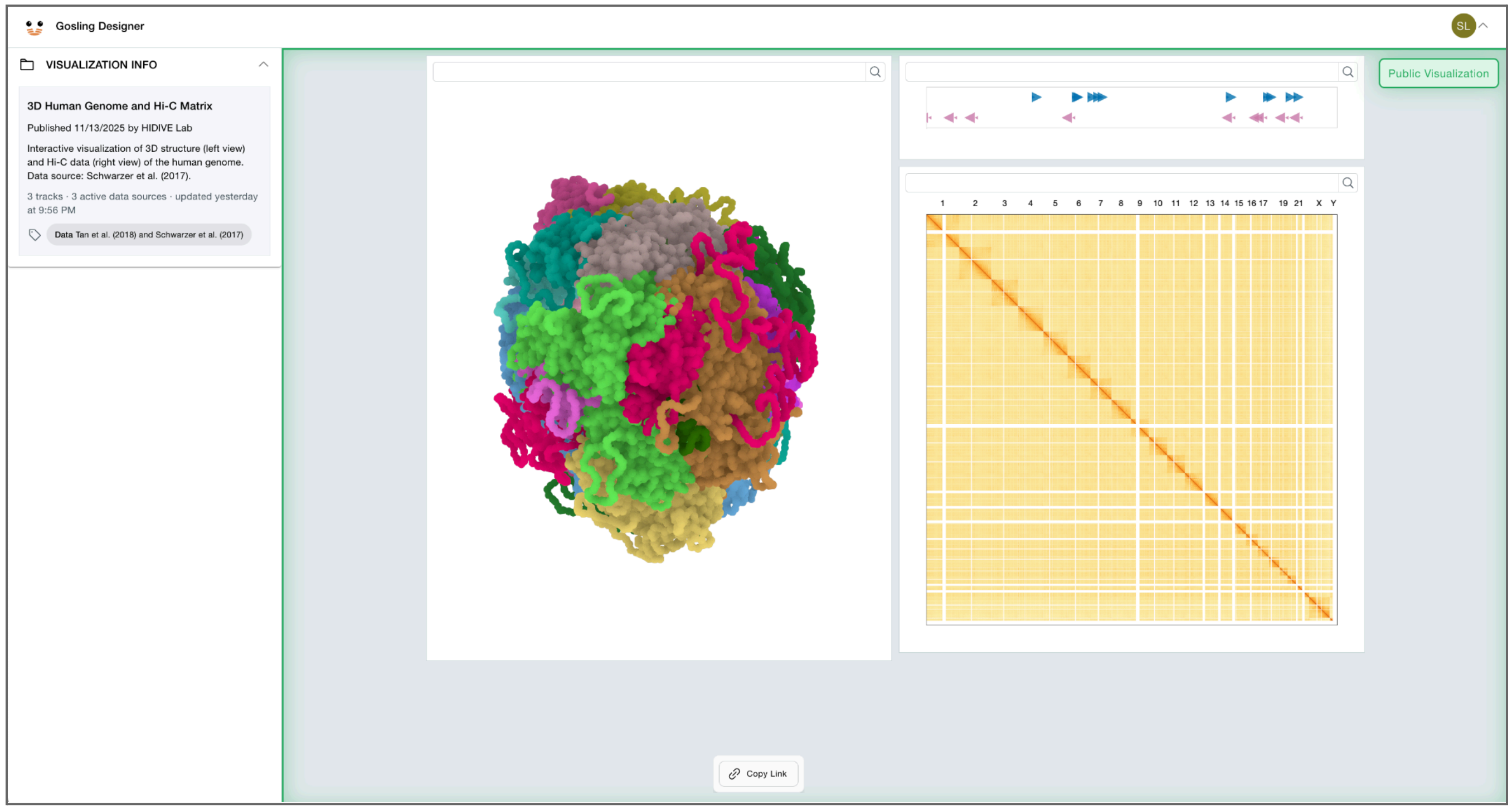

The user interface for a public visualization displays the title, author, description, and other metadata of the visualization (left). The published visualization shown on the right is highlighted with a green outline, indicating that the visualization is accessible by any users. At the bottom, the user can copy a link for this page for sharing with others (e.g., collaborators).

## Sharing Workspaces

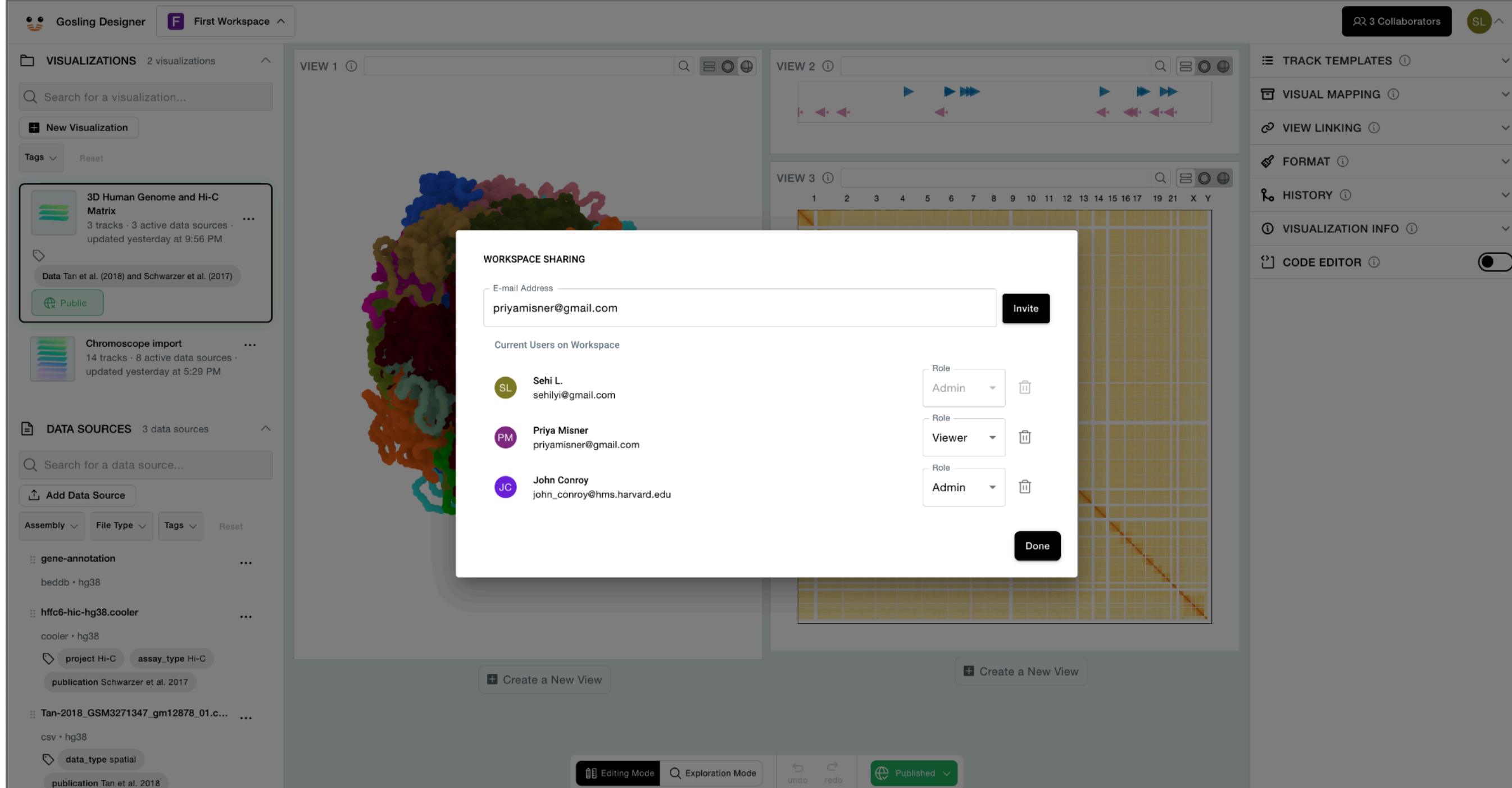

The user can share a workspace to others by clicking the Collaborators button on the top right corner in the header. This view shows the list of current collaborators and their roles (including viewers, editors, and admins). The user can change the role or remove the existing collaborators. Also, at the top of this view, the user can add new collaborators by entering their email addresses.

# Adding Data Sources

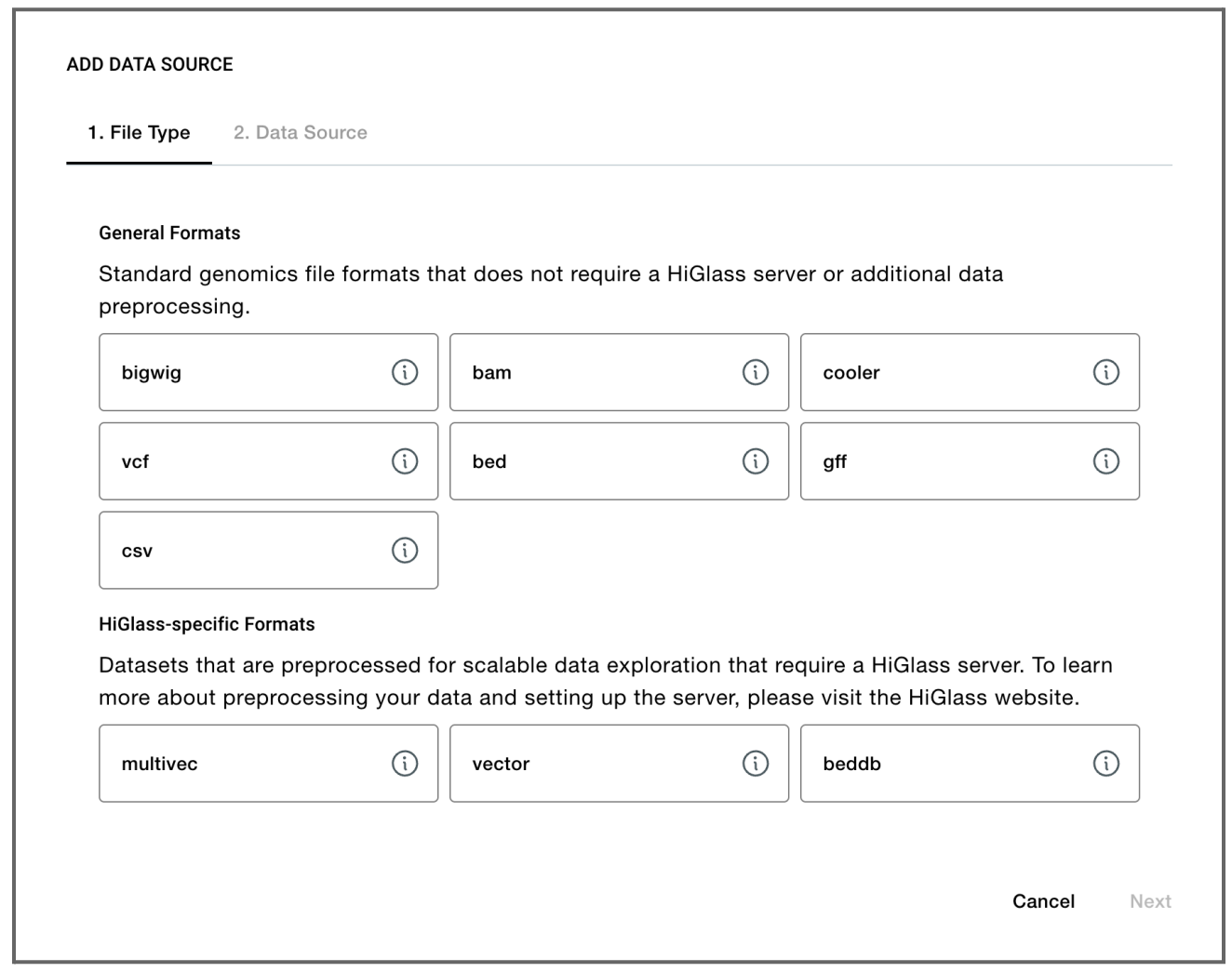

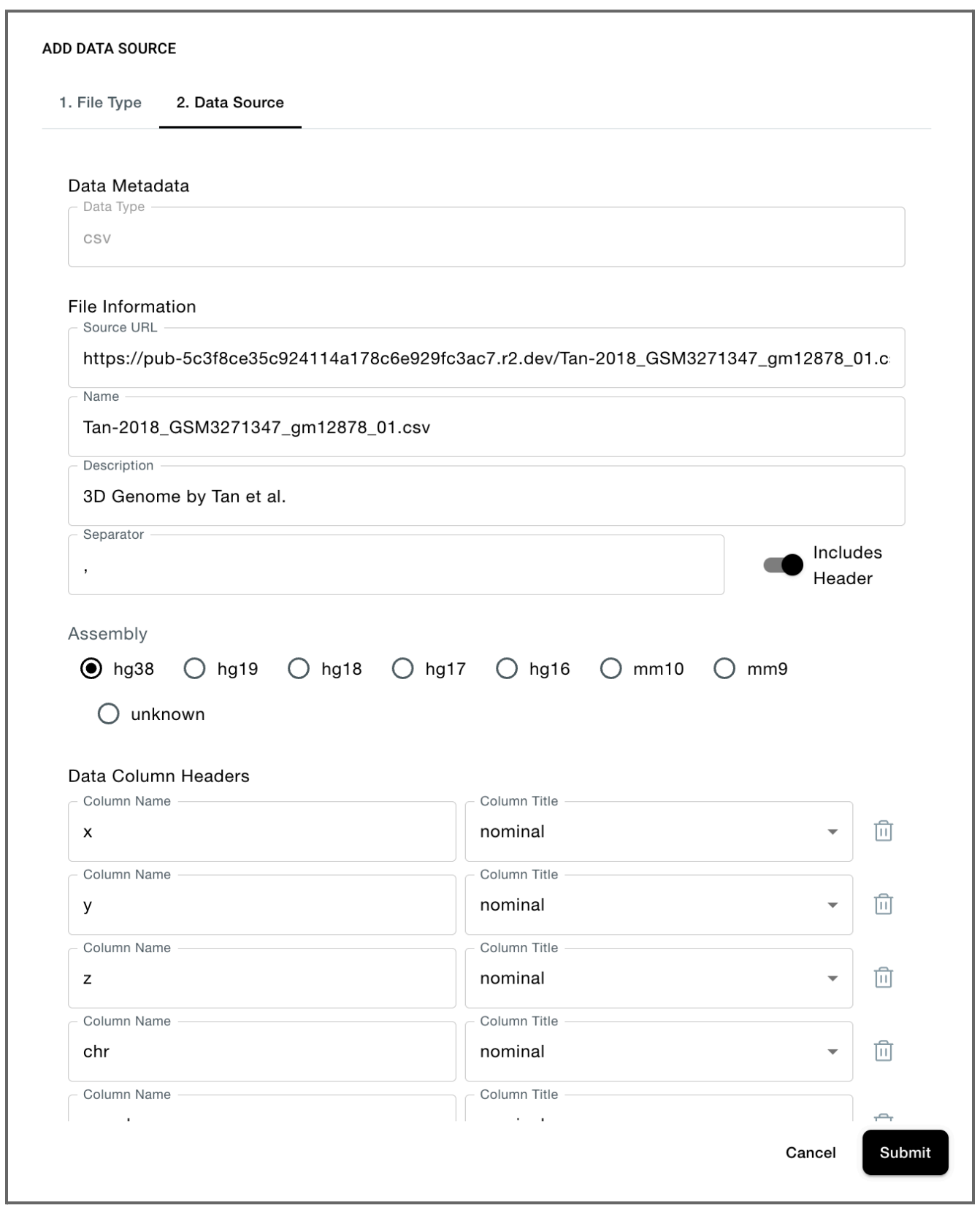

The user can add data sources by clicking the Add Data Source button in the Data Sources panel of the main user interface. This view contains two tabs for (1) selecting a file type and then (2) providing its metadata, such as the URL of a public file, name of the file, description, key-value pair tags, genome assembly, and other information specific to a given file type. Once the user submits the form, the new data source appears in the Data Sources panel and becomes available for visualization.

## Visualization Editing Panel: Track Templates

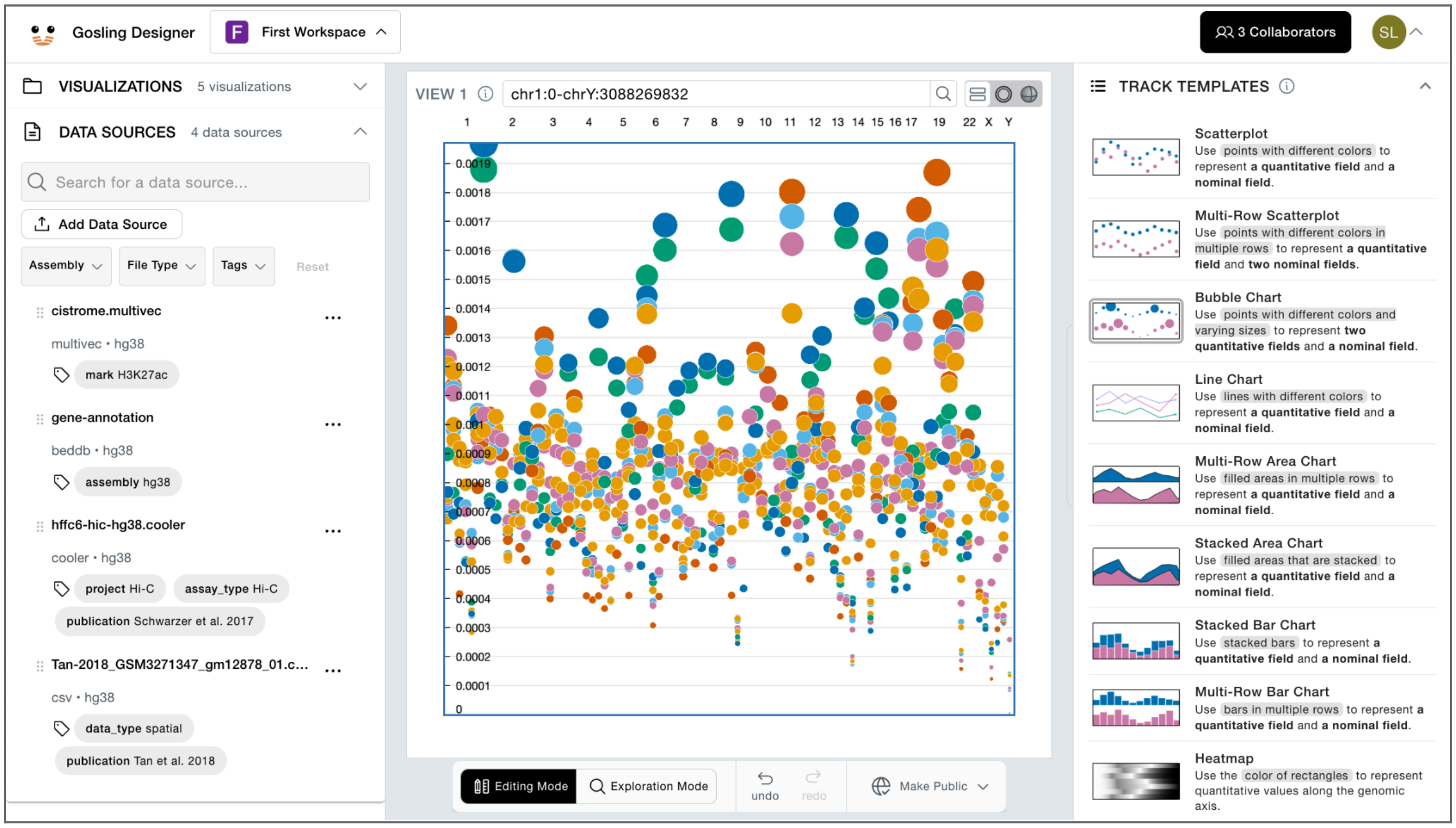

The Track Templates subpanel displays the available track templates for a selected data source, which is chosen by clicking on a track in the visualization canvas. In this example, a BigWig-based track in VIEW 1 has been selected. Based on the data type, Gosling Designer provides a list of available tracks that the user can choose. This enables the user to quickly explore the options with thumbnails and textual descriptions and switch between visualization types.

In this figure, the grey outline in the panel indicates that the Bubble Chart is selected. Hovering over an option previews the result in the visualization canvas (e.g., hovering Line Chart results in visualizing a line chart until after the mouse is moved out, helping the user to quickly see the result).

## Visualization Editing Panel: Visual Mapping

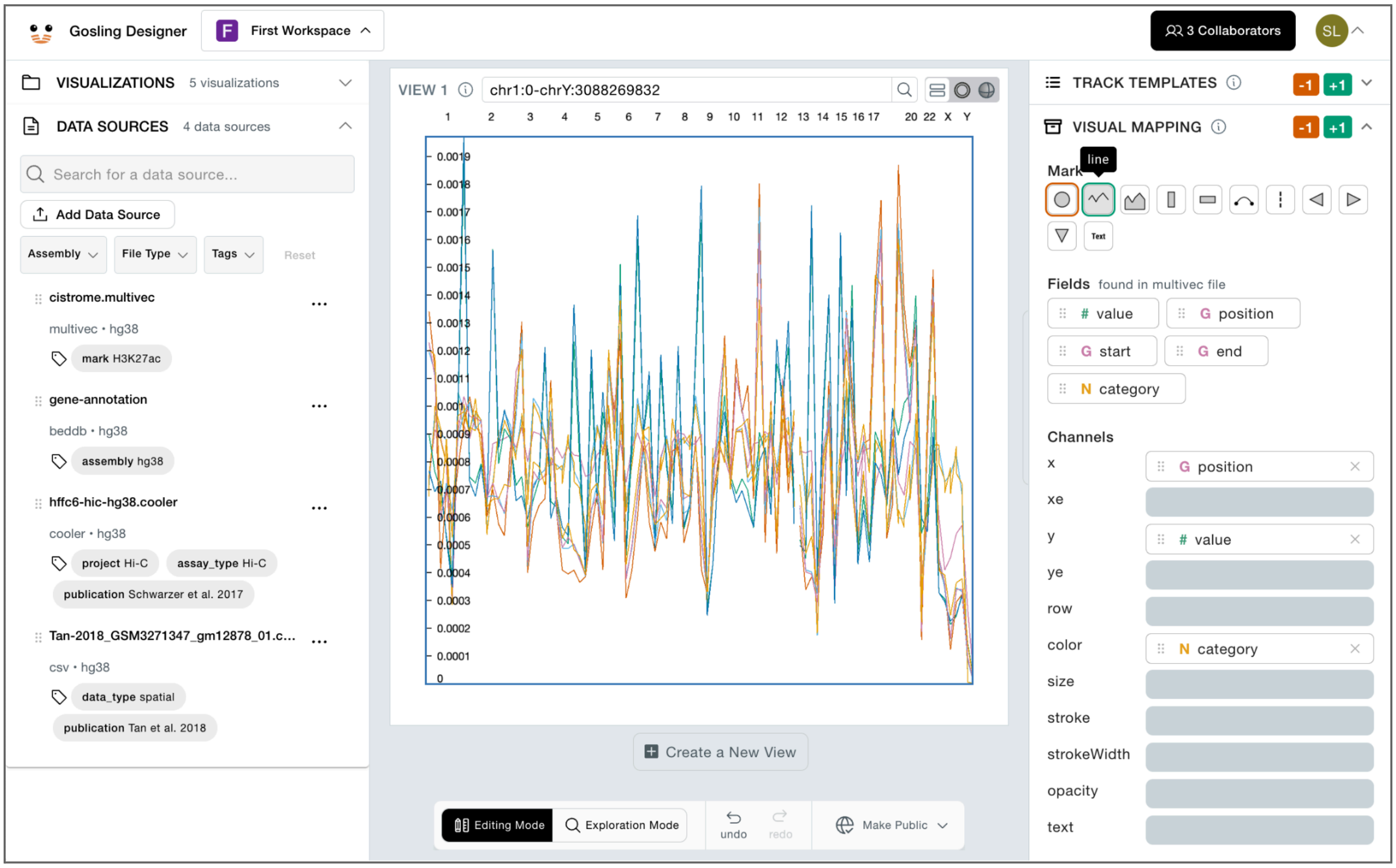

The Visual Mapping subpanel shows available mark types (e.g., points, lines, bars), data fields (i.e., column names available in the data source), and visual channels (e.g., color, size, text) for a selected track. The **G**, **#**, and **N** represent genomic coordinates, quantitative, and nominal values, respectively. Data fields can be dragged and dropped onto visual channels to adjust the visual encoding. For example, dragging a quantitative field to the color channel results in using color to represent quantitative values in the visualization.

If multiple tracks are overlaid, they are shown as tabs in this panel (e.g., two tabs with titles of "Track 1" and "Track 2" when two tracks are overlaid). For example, this will help overlaying a brush on top of a track, which can then be linked with another view (using the View Linking subpanel).

In this figure, the visualization canvas shows a preview for using a line mark (instead of the previously selected point mark) by hovering over the line mark icon.

## Visualization Editing Panel: View Linking

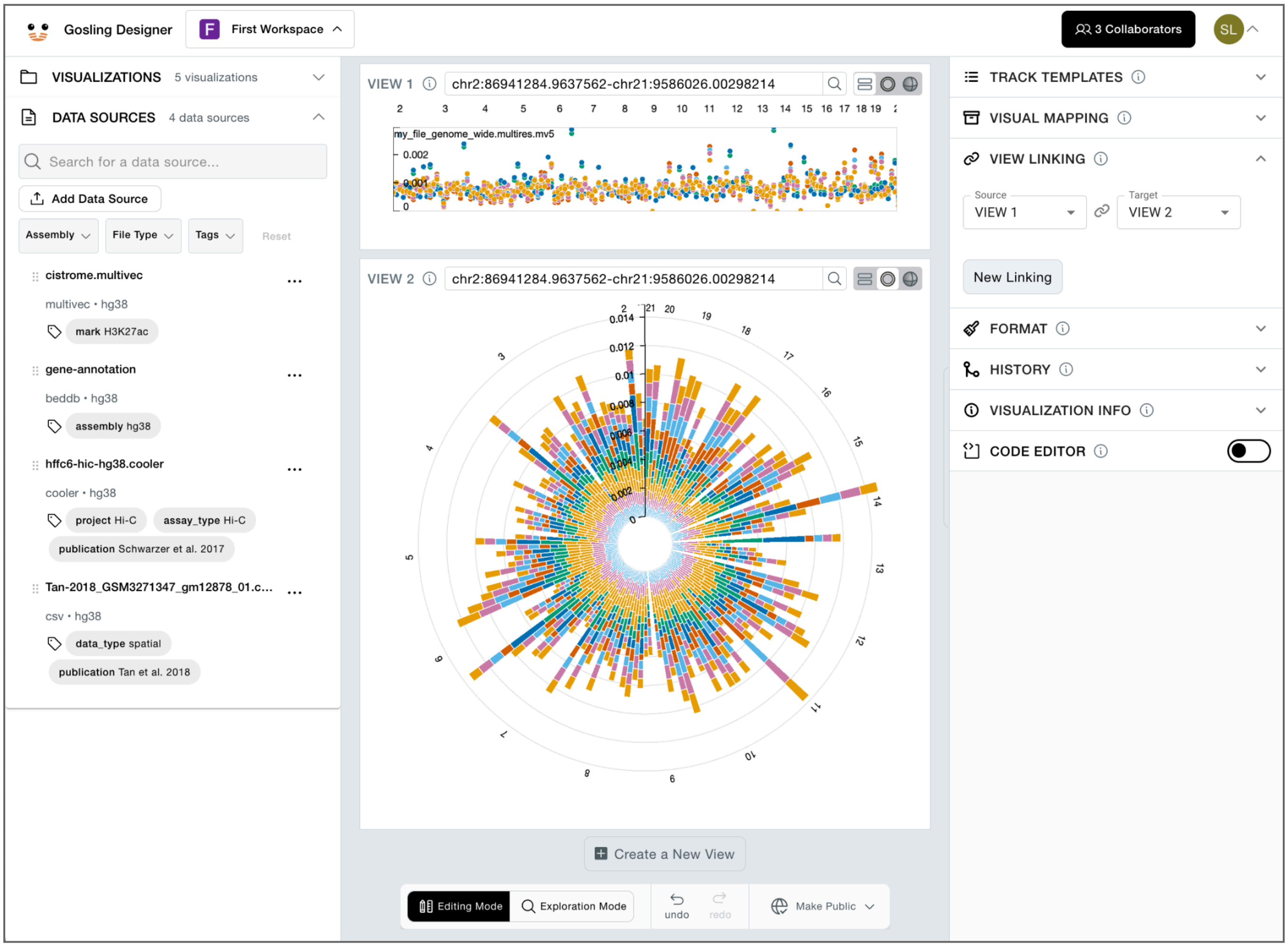

The View Linking subpanel enables the user to link zooming and panning between two or more views. This can be achieved by specifying a pair of views to be linked. If more than two views need to be linked, an additional link can be created by specifying each pair of views to be linked (e.g., linking View 1 and View 2 and then View 2 and View 3).

In this figure, two views in different layouts (linear on the top and circular on the bottom) are linked using the View Linking subpanel.

## Visualization Editing Panel: Format

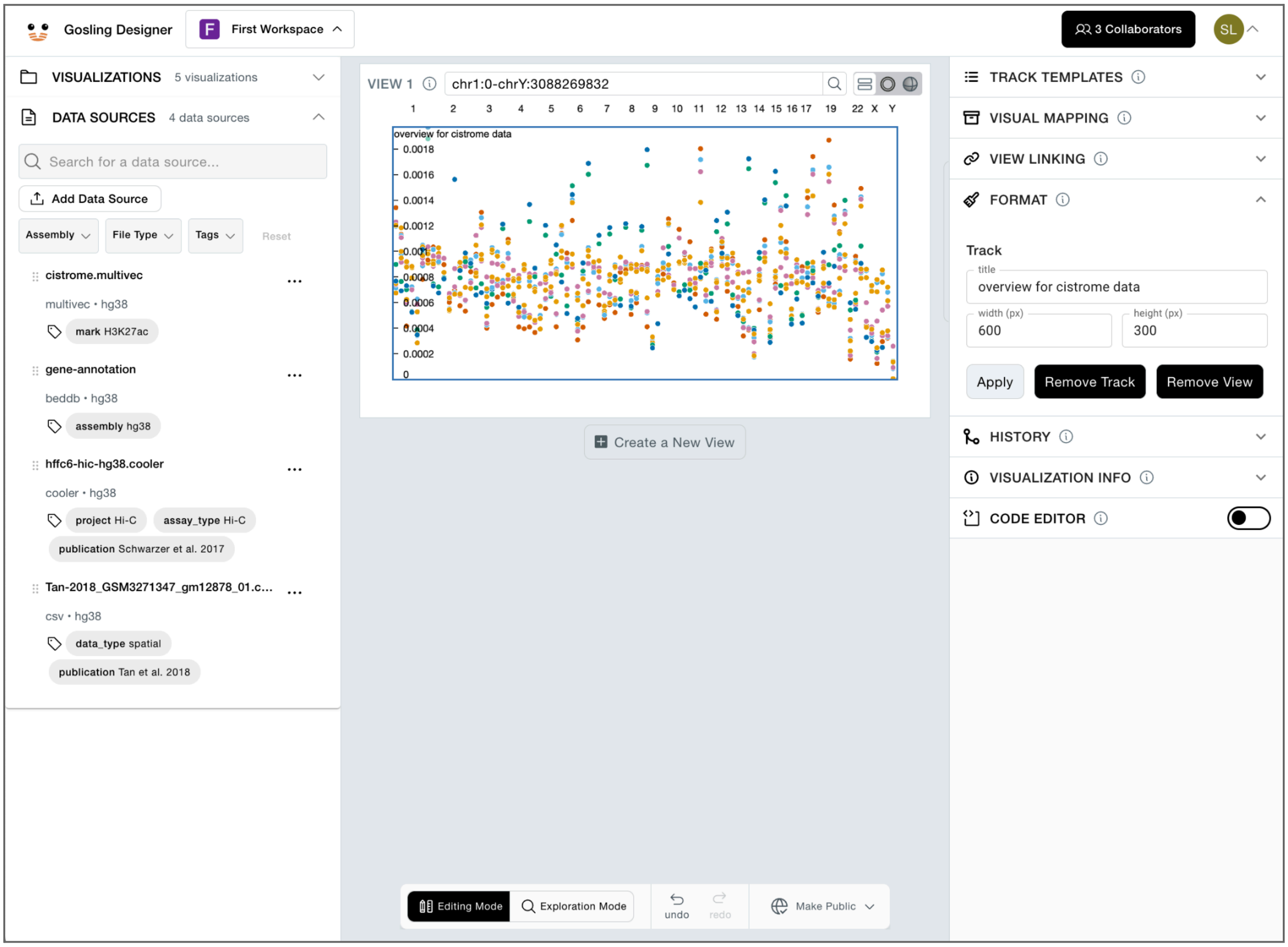

The Format subpanel lets the user configure visual properties of a track or view, such as the track title, width, and height. The selected track and or a corresponding view can be removed by clicking buttons in this panel.

## Visualization Editing Panel: History

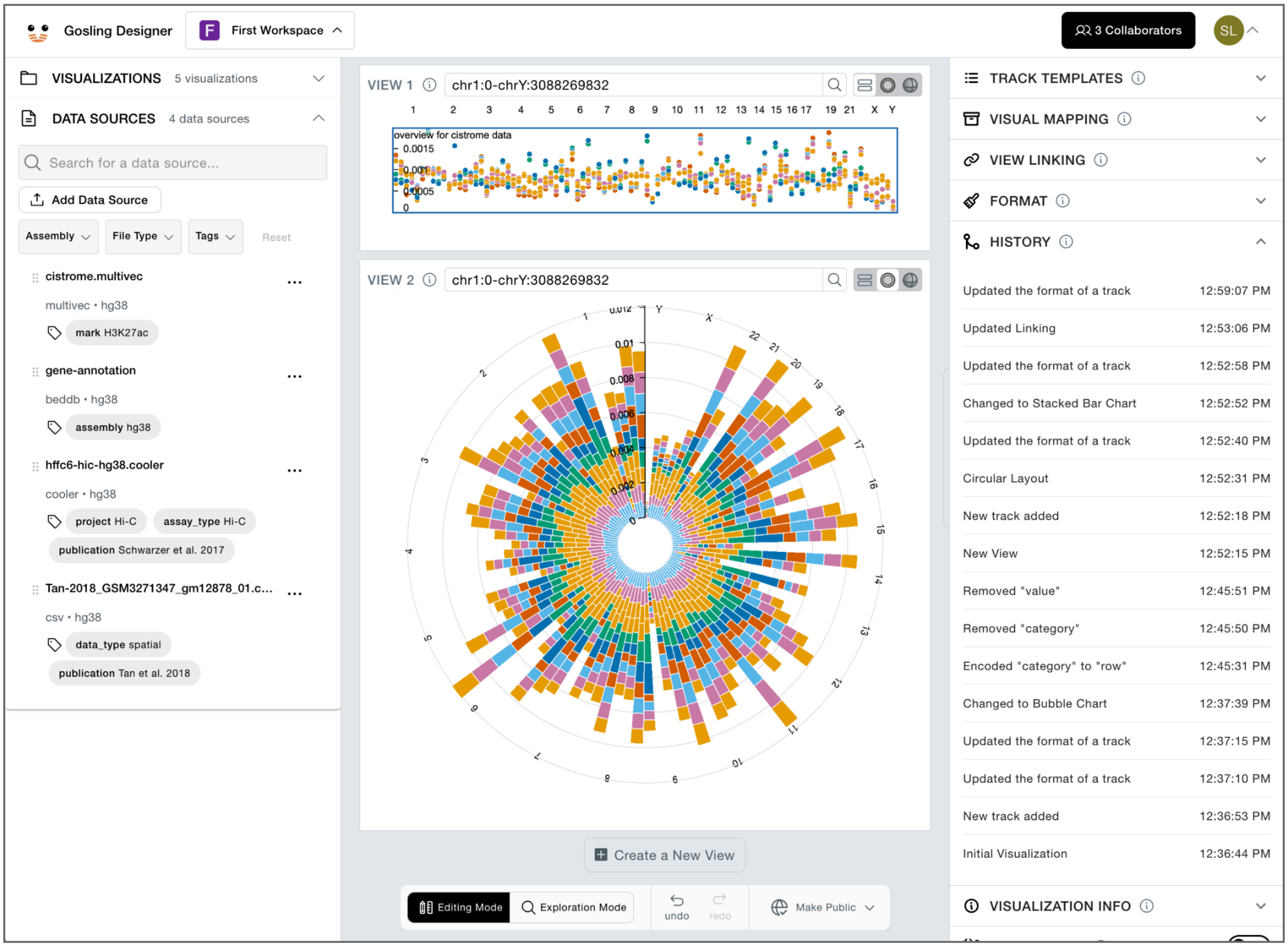

The History subpanel lists all editing actions taken in the session, along with their timestamps. The user can preview the previous state of the visualization by hovering over an item in this list. Clicking on an item allows the user to revert to the corresponding state.

## Visualization Editing Panel: Visualization Info

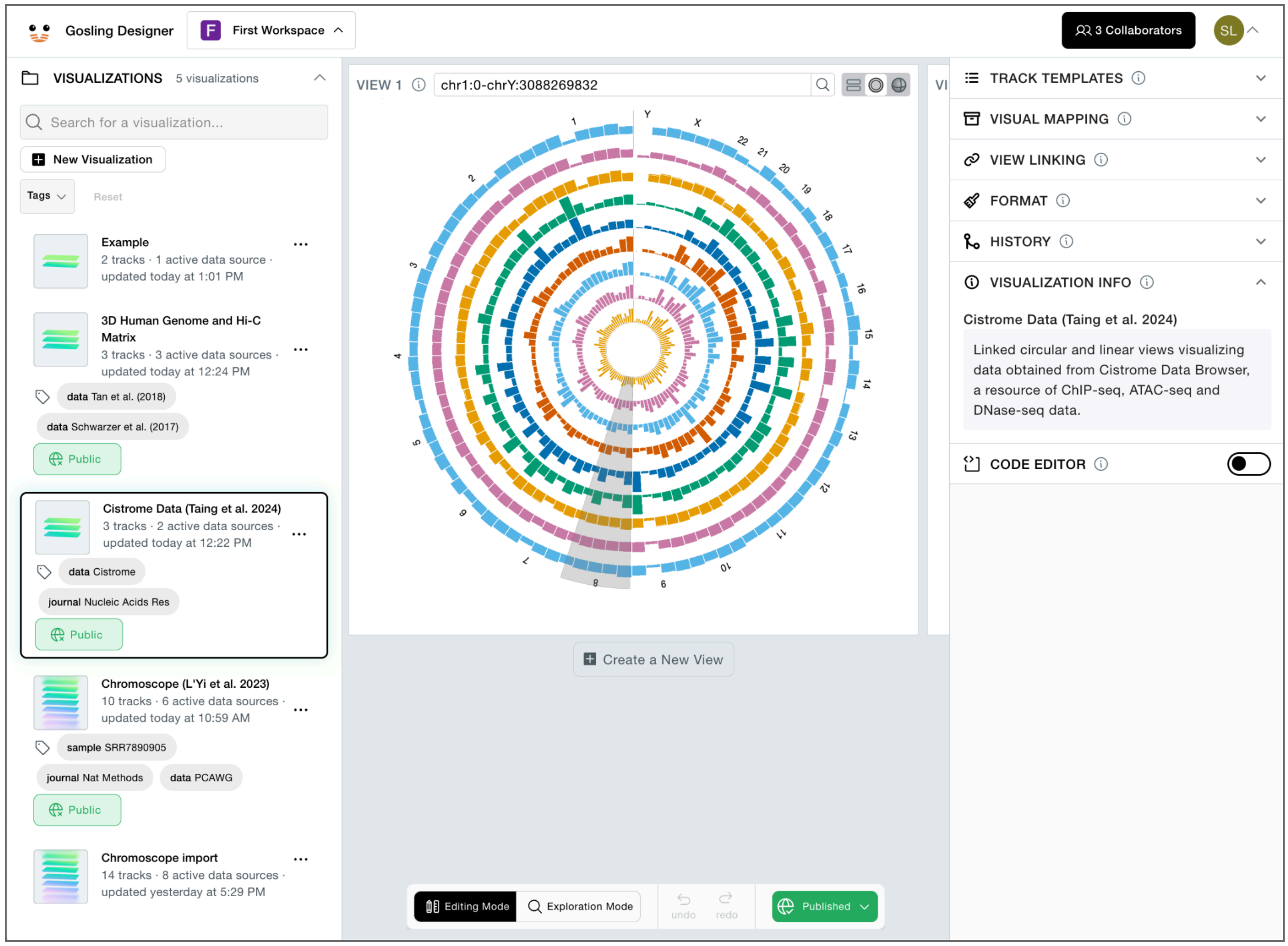

The Visualization Info subpanel shows the title and description of the currently selected visualization. This information can be edited in the Visualizations Panel on the left by selecting the corresponding visualization and clicking the icon with three dots.

## Visualization Editing Panel: Code Editor

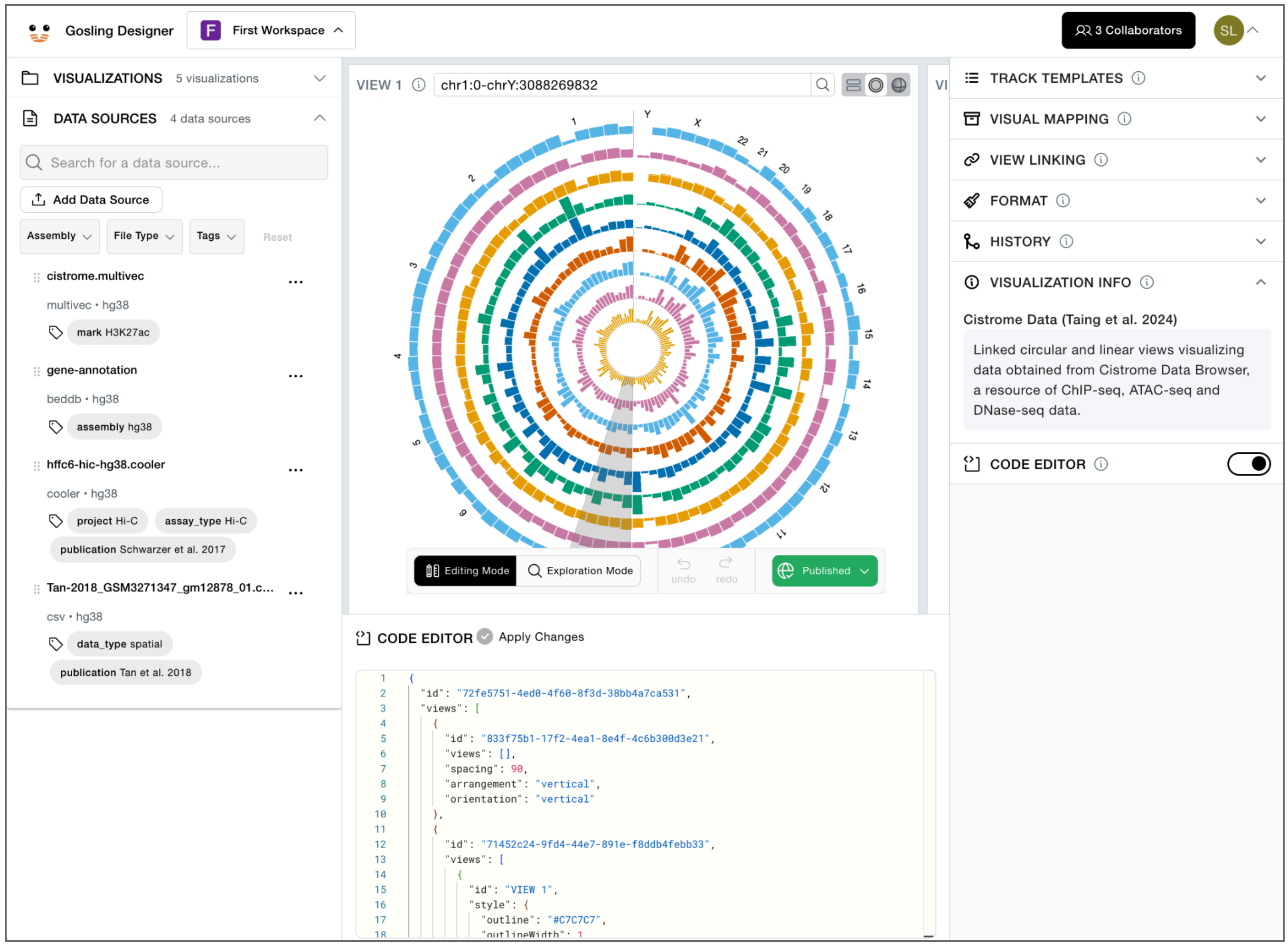

The user can toggle Code Editor, which appears on the bottom of the main user interface. In this view, the user can browse the specification for creating the corresponding visualization and edit it to make further fine-grained customization of the visualization.

## 6. Visualization Examples

We present four visualization examples created with Gosling Designer. For each example, we briefly describe the type of visualization using the taxonomy of Nusrat et al. (2018)[5] for *views*, *scales*, and *foci*, and specify the data being visualized.

### Corces et al. (2020) Single-cell Epigenomic Analysis

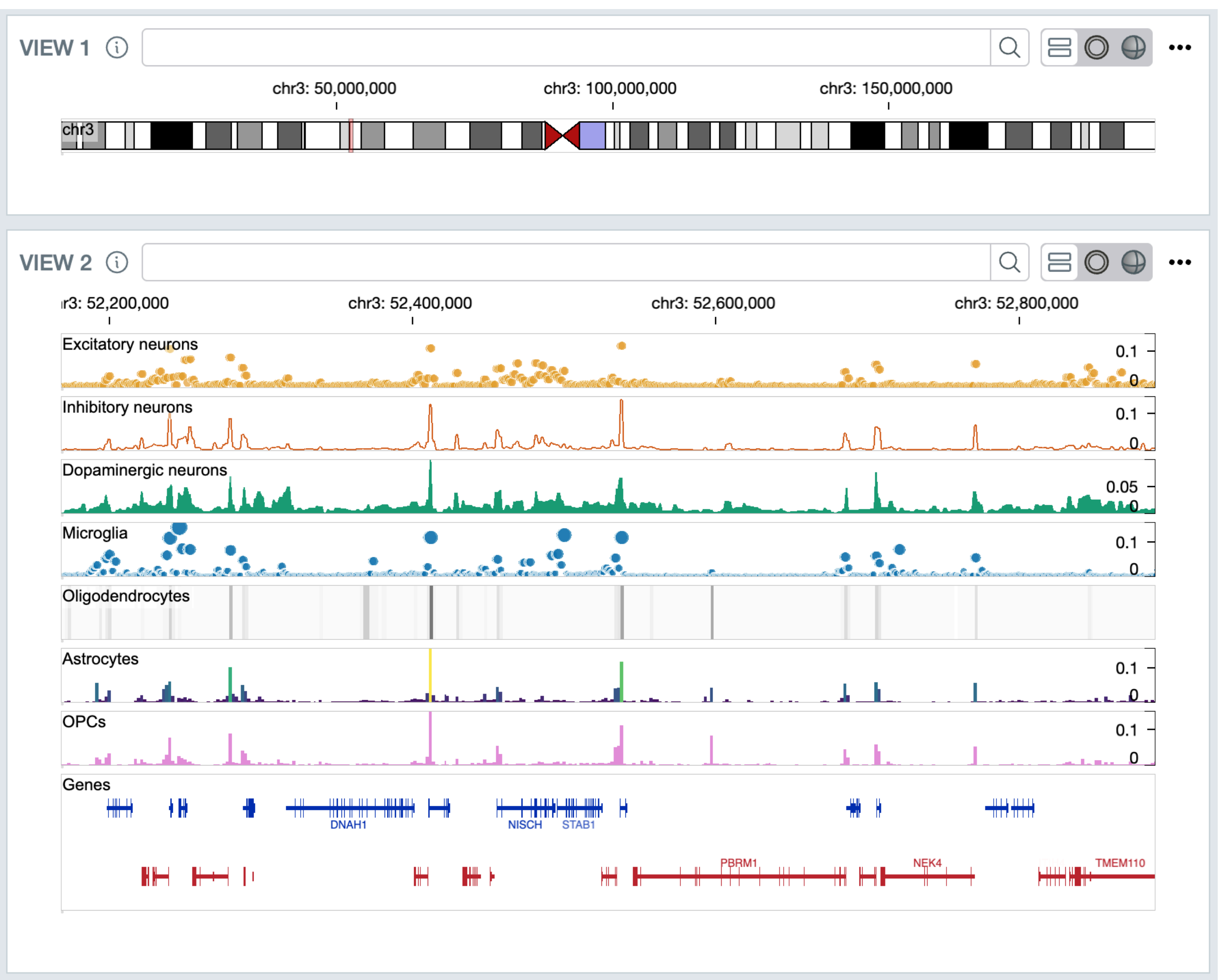

Interactive visualization of the single-cell epigenomic analysis shown in Fig 4a. of Corces et al. (2020)[6]. The visualization is configured as two linked views, at two scales, and with one focus region, in an overview-and-details display. View 1 provides an overview of chromosome 3, showing cytoband information in a single track. View 2 presents detailed epigenetic data for the ITIH1 gene locus (chr3:52,168,000–52,890,000) in seven scATAC–seq tracks, where chromatin accessibility signals for different cell types are represented with different marks and channel combinations (in order: point, line, area, bubble, heatmap, colored bar, and bar), and one gene annotation track, respectively. Data sources: Corces et al. (2020)[6].

## Comparative Matrices

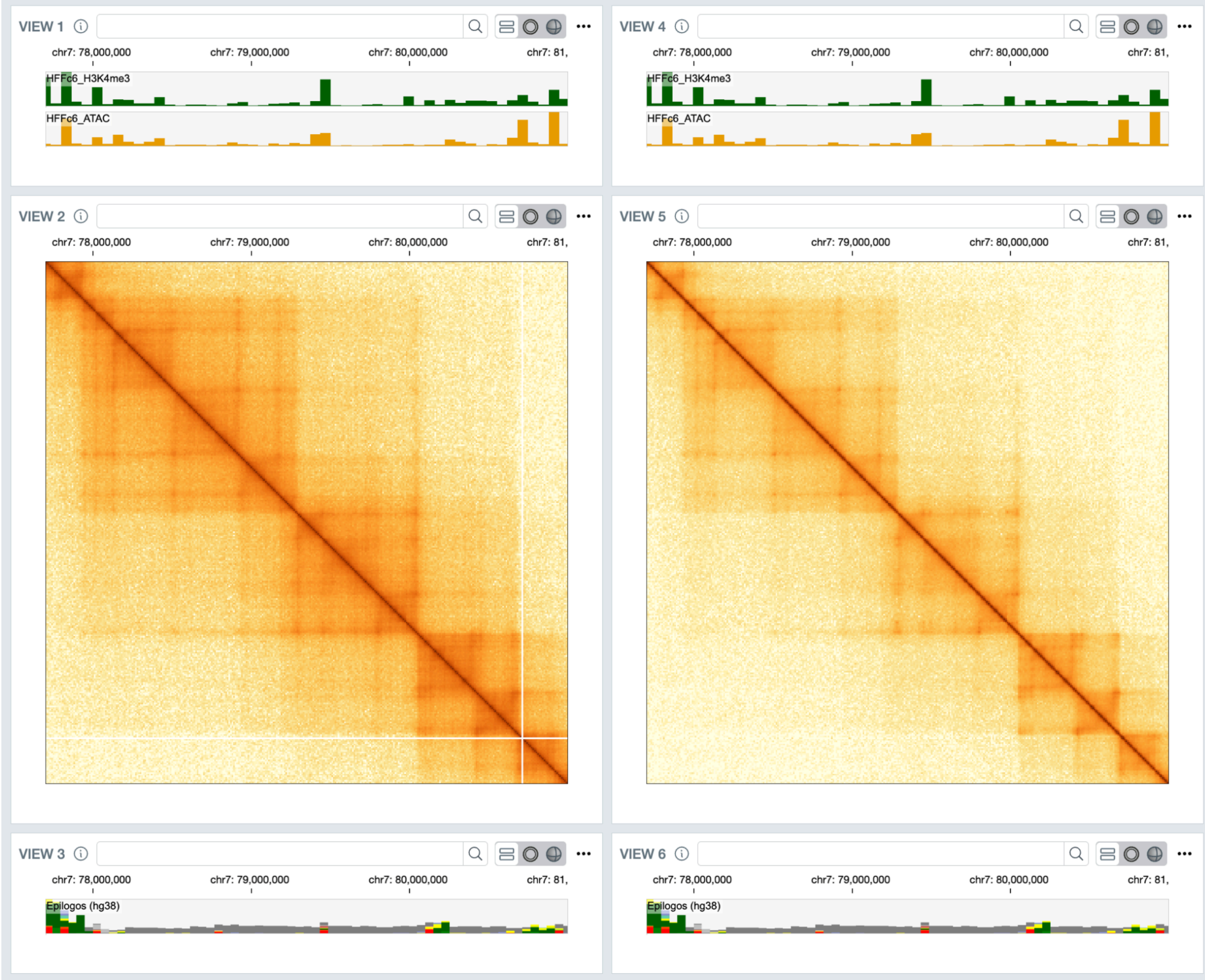

Interactive visualization consisting of Hi-C and Micro-C heatmap matrices of HFF c6 cells, where the heatmap encodes the interaction frequencies with color, each with two top-aligned epigenomic signal tracks using the bar mark, and bottom-aligned epilogos tracks using stacked bars. The composition shows multiple linked linear and orthogonal views, on one scale, and has foci in multiple dimensions (i.e., 1D for the aligned linear tracks and 2D for the matrices). Importantly, the individual views of the visualization do support multiscale exploration by using zooming and panning interactions. Data source: Krietenstein et al. 2020[7].

## 3D Human Genome and Hi-C Matrix

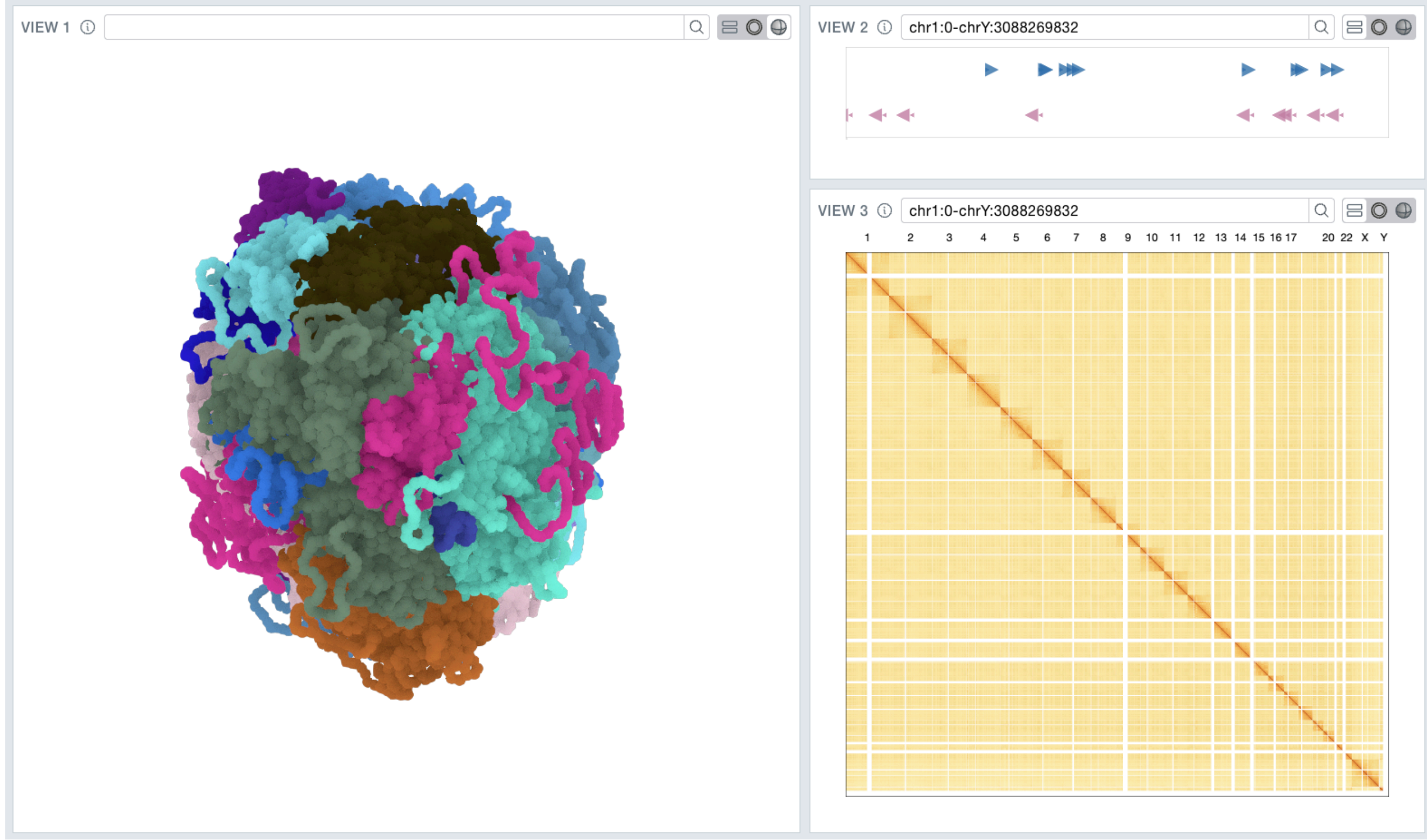

Interactive visualization of 3D structure (left view) and Hi-C data with gene annotations (right views) of the human genome. The left view encodes the different chromosomes in 3D with different colors. The right matrix view uses orthogonal axes to encode interaction frequencies with color. The gene annotations are shown above the matrix view using the standard genome browser track. The visualization is configured as two independent views, on one scale, and with two foci (one being 3D and the other 2D). Data sources: Tan et al. (2018)[8]; Krietenstein et al. 2020[7].

## Coverage, Sequence, and Pileup Plots

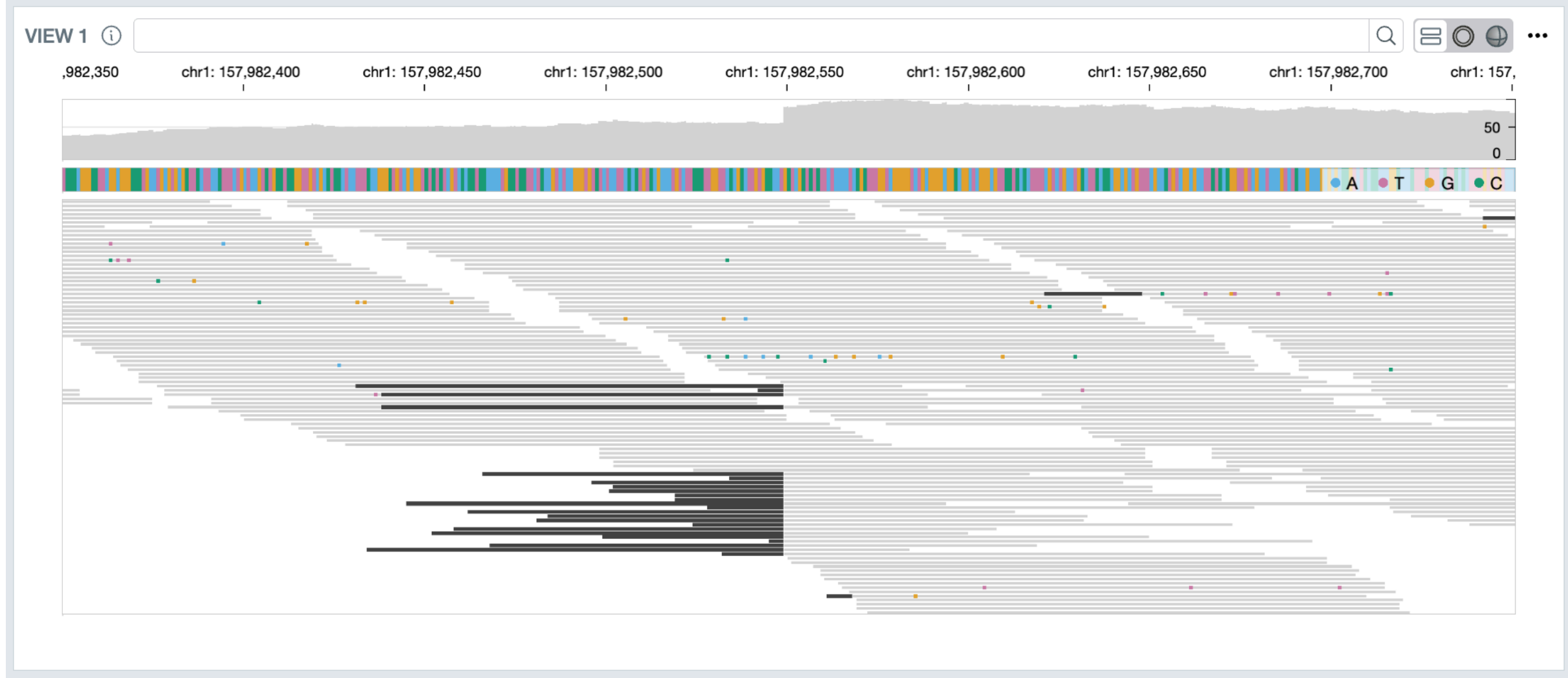

Interactive visualization of read coverage as an area track (top), a multiscale sequence track using color coded bars for the nucleotide types (middle), and sequencing read pileups including identified variants using bar and point marks (bottom). This is a single view, single scale and single focus visualization. Zooming and panning allows multiscale exploration. Upon zooming out, the sequence track transitions into a stacked bar chart, summarizing the frequency of bases. Visualization and data source: L'Yi et al. (2023)[4]. Sample ID: SRR7890905[10] (tumor T1 sample).

Linked Linear and Circular Views

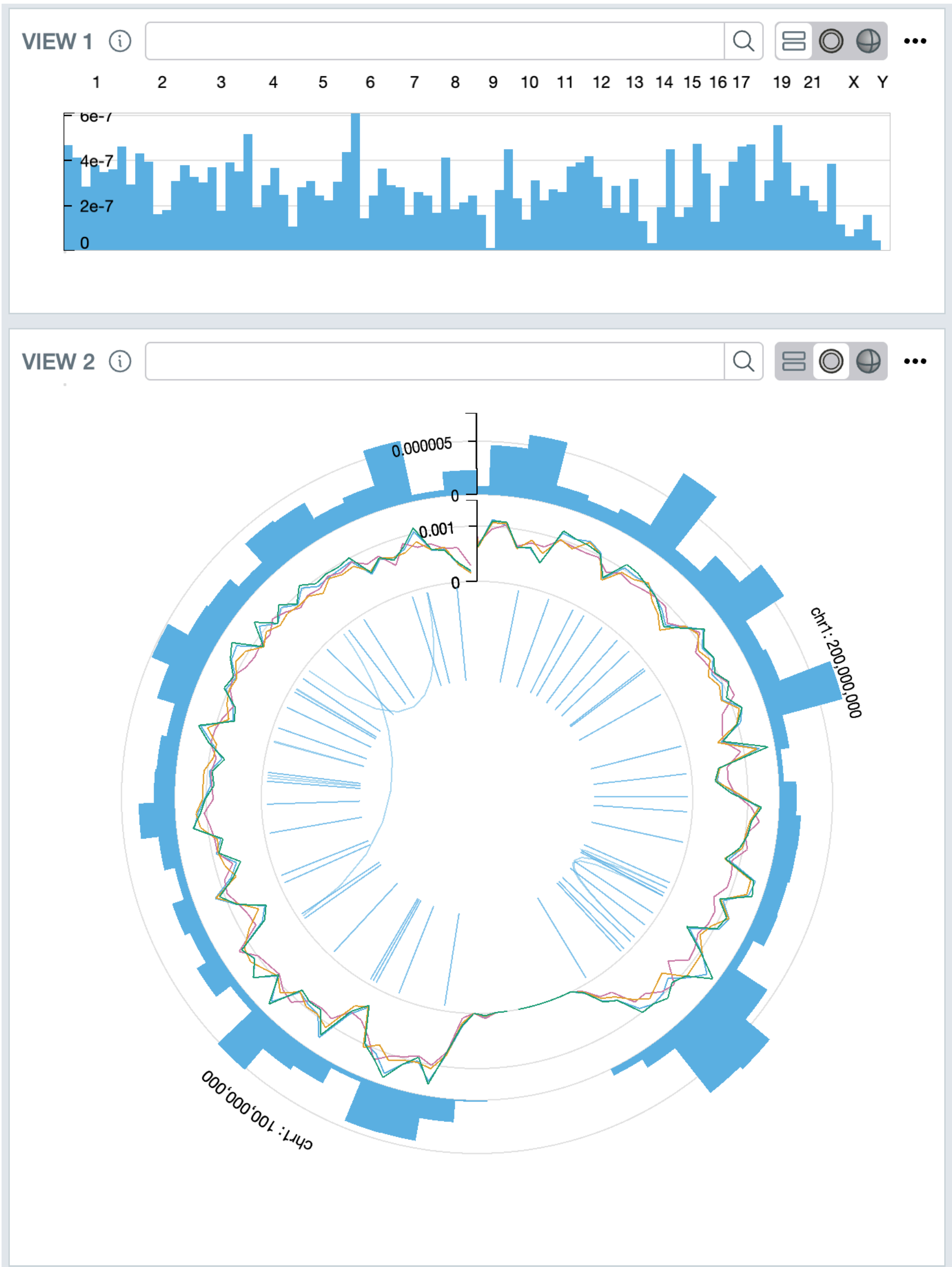

Two linked interactive views of genome-mapped data in linear and circular layout, showing multiple scales and one focus region. The linear view shows a genome-wide overview of ChIP-seq data with the bar mark as a distribution. The circular view shows three stacked tracks with different data types for chromosome 1: ChIP-seq peaks as bars, a track with multiple overlaid line plots for multivec data, and structural variants as arcs in the innermost track. Data source: Schwarzer et al. (2017)[9]; Zheng et al. (2019)[11].

# References

1. L'Yi, S., Wang, Q., Lekschas, F. & Gehlenborg, N. Gosling: A grammar-based toolkit for scalable and interactive genomics data visualization. *IEEE Trans. Vis. Comput. Graph.* **28**, 140–150 (2022).
2. Kerpedjiev, P. *et al.* HiGlass: web-based visual exploration and analysis of genome interaction maps. *Genome Biol.* **19**, 125 (2018).
3. Manz, T., L'Yi, S. & Gehlenborg, N. Gos: a declarative library for interactive genomics visualization in Python. *Bioinformatics* **39**, (2023).
4. L'Yi, S. *et al.* Chromoscope: interactive multiscale visualization for structural variation in human genomes. *Nat. Methods* **20**, 1834–1835 (2023).
5. Nusrat, S., Harbig, T. & Gehlenborg, N. Tasks, Techniques, and Tools for Genomic Data Visualization. *Comput. Graph. Forum* **38**, 781–805 (2019).
6. Corces, M. R. *et al.* Single-cell epigenomic analyses implicate candidate causal variants at inherited risk loci for Alzheimer's and Parkinson's diseases. *Nature Genetics* vol. 52 1158–1168 Preprint at https://doi.org/10.1038/s41588-020-00721-x (2020).
7. Krietenstein, N. *et al.* Ultrastructural Details of Mammalian Chromosome Architecture. *Mol. Cell* **78**, 554–565.e7 (2020).
8. Tan, L., Xing, D., Chang, C.-H., Li, H. & Xie, X. S. Three-dimensional genome structures of single diploid human cells. *Science* **361**, 924–928 (2018).
9. Schwarzer, W. *et al.* Two independent modes of chromatin organization revealed by cohesin removal. *Nature* **551**, 51–56 (2017).
10. Fang, L. T. *et al.* Establishing community reference samples, data and call sets for benchmarking cancer mutation detection using whole-genome sequencing. *Nat. Biotechnol.* **39**, 1151–1160 (2021).
11. Zheng, R. *et al.* Cistrome Data Browser: expanded datasets and new tools for gene regulatory analysis. *Nucleic Acids Res.* **47**, D729–D735 (2019).